\documentclass[reqno]{amsart}
\usepackage[centertags]{amsmath}
\usepackage{amsfonts,amssymb, amsthm}
\usepackage{hyperref}
\usepackage{comment}

\usepackage{cite,graphicx,color,ulem}
\usepackage{fourier}
\usepackage[margin=1.5in]{geometry}
\usepackage{enumitem}

\newtheorem{theorem}{Theorem}
\newtheorem{assumption}{Assumption}

\def\EE{\mathbb{E}}
\def\PP{\mathbb{P}}
\def\RR{\mathbb{R}}
\newcommand{\mD}{\mathcal{D}}
\newcommand{\mF}{F}

\DeclareMathOperator{\erf}{erf}
\DeclareMathOperator{\erfc}{erfc}
\DeclareRobustCommand{\argmin}{\operatorname*{argmin}}

\def\EE{\mathbb{E}}\def\PP{\mathbb{P}}
\def\NN{\mathbb{N}}\def\RR{\mathbb{R}}\def\ZZ{\mathbb{Z}}

\def\<{\langle} \def\>{\rangle}

\begin{document}

\title{Extreme event quantification in dynamical systems with random
  components}

\date{\today}

\author{Giovanni Dematteis}
\address{Dipartimento di Scienze Matematiche, Politecnico di
  Torino, Corso Duca degli Abruzzi 24, I-10129 Torino, Italy}

\author{Tobias Grafke}
\address{Mathematics Institute, University of Warwick, Coventry CV4
  7AL, United Kingdom}

\author{Eric Vanden-Eijnden}
\address{Courant Institute, New York University, 251 Mercer
  Street, New York, NY 10012, USA}

\begin{abstract}
  A central problem in uncertainty quantification is how to
  characterize the impact that our incomplete knowledge about models
  has on the predictions we make from them. This question naturally
  lends itself to a probabilistic formulation, by making the unknown
  model parameters random with given statistics. Here this approach is
  used in concert with tools from large deviation theory (LDT) and
  optimal control to estimate the probability that some observables in
  a dynamical system go above a large threshold after some time, given
  the prior statistical information about the system's parameters
  and/or its initial conditions. Specifically, it is established under
  which conditions such extreme events occur in a predictable way, as
  the minimizer of the LDT action functional. It is also shown how
  this minimization can be numerically performed in an efficient way
  using tools from optimal control. These findings are illustrated on
  the examples of a rod with random elasticity pulled by a
  time-dependent force, and the nonlinear Schr\"odinger equation
  (NLSE) with random initial conditions.
\end{abstract}
\maketitle

\section{Introduction}
\label{sec:intro}

The governing equations we use to model complex phenomena are often
approximate. For example, we may not know exactly the initial and/or
boundary conditions necessary to integrate these equations. Other
parameters entering these equations can also be uncertain, either
because we are not sure of the model itself or because these
parameters may vary from situations to situations in a way that is
difficult to predict in detail. The question then becomes whether we
can quantify how our imperfect knowledge of the system's parameters
impact its behavior. This question lends itself naturally to a
probabilistic formulation. Consider for example the case of a
dynamical system whose state at time~$t$ can be specified by some
$u(t)$ which can be a vector or a field and satisfies
\begin{equation}
  \label{eq:detdyn}
  \partial_t u = b(u,\vartheta), \qquad u(t=0) = u_0(\vartheta).
\end{equation}
Here $b(u,\vartheta)$ is a given vector field and $\vartheta$ denotes
the set of parameters we are uncertain of. Assuming that these
parameters take value in some set $\Omega$, which can again be finite
or infinite dimensional, it is then natural to equip $\Omega$ with a
probability measure $\mu$ to quantify our uncertainty. This makes
$\vartheta$ random, and therefore the solution to~\eqref{eq:detdyn}
becomes a stochastic process. Denoting it by $u(\cdot,\vartheta)$, we
can ask questions about the statistics of this process. For example,
if $f(u)$ is a scalar valued observable, we can define
\begin{equation}
  \label{eq:expectations}
  P_T(z) \equiv 
  \PP \left(f(u(T,\vartheta))\ge z\right), \qquad z\in \RR\,,
\end{equation}
where $\PP$ denotes the probability over~$\mu$ and $T>0$ is some
observation time. The probability \eqref{eq:expectations} is useful
e.g.~in the context of certification problem where, given $z\in \RR$
and $\epsilon>0$ (typically $z$ large and $\epsilon$ small), we wish
to verify that $P_T(z)\le \epsilon$.  Other quantities of interest
include
\begin{equation}
  \label{eq:3}
  \PP \left(\int_0^T f(u(t,\vartheta)) dt\ge z\right),  \qquad
  \PP \left(\sup_{0\le t \le T}  f(u(t,\vartheta)) \ge z\right),
  \qquad \text{etc.}
\end{equation}

The numerical estimation of~\eqref{eq:expectations} or~\eqref{eq:3}
can be performed by Monte Carlo sampling methods: generate $N$
independent realizations of $\vartheta$, for each evaluate
$f(u(T,\vartheta))$ via integration of~\eqref{eq:detdyn}, and compute
the fraction of these realizations for which $f(u(T,\vartheta))\ge
z$. As $N\to\infty$, this fraction will converge to $P_T(z)$. This
direct approach is not effective when $P_T(z)$ is small, however,
since the relative error of the estimator just described is
$\sqrt{(1-P_T(z))/(NP_T(z))} \sim 1/\sqrt{NP_T(z)}$. This means that
in order to get an estimate accurate to order $\delta \ll1$, we need
to use $N=O\left(\delta^{-2}P^{-1}_T(z)\right)$ samples, which can
become prohibitively expensive as $P_T(z)$ gets smaller. This is
problematic since it excludes from consideration events that are rare
but may nonetheless have dramatic consequences. Similar issues arise
if we replace~\eqref{eq:detdyn} by some time independent equation like
\begin{equation}
  \label{eq:detndyn}
  0 = b(u,\vartheta), 
\end{equation}
where $b(\cdot, \vartheta)$ is some function of $u$ and possibly its
derivatives and \eqref{eq:detndyn} is supplemented with boundary
conditions that may also depend on the random
parameter~$\vartheta$. The solution to~\eqref{eq:detndyn} defines a
complicated map $u(\vartheta)$, and given a scalar valued observable
$f(u)$, the estimation of
\begin{equation}
  \label{eq:expectationsndyn}
  \PP \left(f(u(\vartheta))\ge z\right), \qquad z\in \RR
\end{equation}
will again be challenging when this probability is small, i.e.~when
the event $f(u(\vartheta))\ge z $ is rare.

In these situations alternative methods such as those proposed e.g.
in~\cite{glasserman1999multilevel,juneja2006rare,%
  cerou2007adaptive,giardina2011simulating,tailleur2007probing,VandenEijnden:2012ef,%
  farazmand-sapsis:2017,ragone2017computation} must be used to
estimate~\eqref{eq:expectations},~\eqref{eq:3},
or~\eqref{eq:expectationsndyn}. The approach we introduce in this
paper builds on earlier results found in~\cite{PNAS2018rogue} and uses
large deviation theory (LDT)~\cite{dembo2010large,varadhan2016large}
as a tool: we show that, if
in~\eqref{eq:expectations} $P_T(z)\to0$ as $z\to\infty$, then under
some additional assumptions we have
\begin{equation}
  \label{eq:2}
  P_T(z) \asymp \exp\left( -\min_{\theta\in \Omega(z)}\, I(\theta)
  \right)  \qquad\text{where} \qquad
  \Omega(z) = \left\{\theta : f(u(T,\theta))\ge z\right\} \subseteq \Omega.
\end{equation}
Here $\asymp$ indicates that the ratio of logarithms of both sides
tends to 1 as $z\to\infty$ and we defined
\begin{equation}
  \label{eq:6bb}
  I(\theta) = \max_{\eta} \left( \<\eta, \theta\> - S(\eta)
  \right) ,
\end{equation}
where $\<\cdot, \cdot\>$ is a suitable inner product on $\Omega$ and
$S(\eta)$ is the cumulant generating function of $\vartheta$:
\begin{equation}
  \label{eq:5bb}
  S(\eta) = \log \EE e^{\<\eta, \vartheta\>} = \log \int_\Omega 
  e^{\<\eta, \theta\>}d\mu(\theta)\,.
\end{equation}
We will also show that the minimizer of $I(\theta)$ in
$\Omega(z)$, i.e.
\begin{equation}
  \label{eq:2gbb}
  \theta^\star(z) = \argmin_{\theta\in \Omega(z)} I(\theta)\,,
\end{equation}
is the point of maximum likelihood in $\Omega(z)$. The most likely way
the event $\left\{f(u(T,\vartheta))\ge z\right\}$ occurs is when
$\vartheta = \theta^\star(z)$. Similar estimates hold for~\eqref{eq:3}
and~\eqref{eq:expectationsndyn} upon straightforward redefinition of
the set $\Omega(z)$ upon which the optimization is performed.

Establishing the large deviation principle (LDP) in~\eqref{eq:2} is
one of the objectives of this paper. As we will see in
Sec.~\ref{sec:LDTmethod}, this can be done by proving that
$\theta^\star(z) $ is a dominating point in $\Omega(z)$, building on
results derived e.g. in~\cite{Borozkov:1965, Ney:1983iz,
  broniatowski1995tauberian,Iltis:2000vp} that provide us with a
framework to justify the saddle-point approximations often used in
physics~\cite{jensen1995saddlepoint,frisch1997extreme}. Eq.~\eqref{eq:2}
is a somewhat unusual LDP however because there is no small (or large)
parameter associated to the random variable $\vartheta$: rather we
play with the variable $z$ being large. More precisely, instead of
scaling $\vartheta$ so that events with a finite $z$ become rare, we
keep $\vartheta$ as is and look at rare events that occur in the tail
of the distribution when $z\gg1$. As a result, the standard approach
developed in~\cite{Borozkov:1965, Ney:1983iz, Iltis:2000vp} must be
adapted. Of course, both viewpoints are equivalent up to some
appropriate rescaling of the variables $\vartheta$ and $z$, but this
rescaling involves the so-called speed of the LDP, which is unknown to
us \textit{a~priori}. The formulation we adopt can be viewed as a way
to estimate this speed.

When~\eqref{eq:2} holds, we can reduce the evaluation of $P_T(z)$ to
the minimization problem in~\eqref{eq:2gbb}, and a second objective
here is to design numerical tools to perform this minimization. As we
will see in Sec.~\ref{sec:optimization}, this can be done by adapting
techniques used in optimal
control~\cite{troltzsch2010optimal,borzi2011computational}.

We will also illustrate these tools on two examples in
Sec. ~\ref{sec:applications}: The first one is a model for an elastic
rod with a random elasticity coefficient. The rod gets pulled from one
end with a given forcing protocol, and the response depends
nonlinearly on the elasticity coefficient. The LDP can be used here to
infer the probability of atypically large extensions of the rod.  The
second application deals with the nonlinear Schr\"odinger equation
(NLSE) in nonlinear fiber optics, in the context of what is known as
integrable turbulence, and study the problem of the onset of rogue
waves out of a bath of random waves taken as initial condition for
NLSE.

\section{Large deviation principle}
\label{sec:LDTmethod}

Here we establish~\eqref{eq:2}, using background material that can be
e.g.~found in~\cite{Borozkov:1965,Ney:1983iz,Iltis:2000vp}.  For
simplicity, we will restrict ourselves to situations where $\vartheta$
is finite dimensional, i.e.~we assume that
$\vartheta \in \Omega \subseteq \RR^M$ with $M\in \NN$. In this case
we can also assume that the inner product $\< \cdot,\cdot\>$ appearing
in~\eqref{eq:6bb} and~\eqref{eq:5bb} is the standard Euclidean inner
product on $\RR^M$. Under appropriate assumptions, the results below
will hold also in the infinite-dimensional set-up, when $\vartheta$ is
a random field, but the arguments to establish them will require
generalization (see e.g.~\cite{Einmahl:1996cg,kuelbs2000large} for
results in infinite dimension). To treat the problems
in~\eqref{eq:expectations},~\eqref{eq:3}
and~\eqref{eq:expectationsndyn} on the same footing we also define the
map $F:\Omega \to\RR$ via
\begin{equation}
  \label{eq:1}
  \begin{aligned}
    &F(\theta) = f(u(T,\theta)), \qquad
    &\text{for~\eqref{eq:expectations}} \\ & F(\theta)
    = \int_0^T f(u(T,\theta))\, dt, \quad\text{or}\quad F(\theta) =
    \sup_{0\le t\le T} f(u(T,\theta)), &\text{for~\eqref{eq:3}}\\
    &F(\theta) = f(u(\theta))
    &\text{for~\eqref{eq:expectationsndyn}}
  \end{aligned}
\end{equation}
so that  we can recast these probabilities into
\begin{equation}
  \label{eq:1b}
  P(z) = \PP(\mF(\vartheta) \ge z ) = \mu(\Omega(z)) \qquad
  \text{where} \qquad \Omega(z)= \{ \theta \, : \, \mF(\theta)\ge
  z\}\,.
\end{equation}
To proceed, we start by making two  assumptions:
\begin{assumption}
  \label{th:as1} The map $F$ is continuously differentiable, and such
  that $|\nabla F(\theta)|\ge K >0$ for all $\theta\in \Omega$.
\end{assumption}
\begin{assumption}
  \label{th:as3} The measure $\mu$ is such that (this
  is~\eqref{eq:5bb})
\begin{equation}
  \label{eq:5b}
  S(\eta) = \log \EE e^{\<\eta, \vartheta\>} = \log \int_\Omega 
  e^{\<\eta, \theta\>}d\mu(\theta)
\end{equation}
exists for all $\eta\in \RR^M$ and defines a differentiable function
$S:\RR^M \to \RR$.
\end{assumption}
\noindent
Ultimately, Assumption~\ref{th:as1} is about the specifics of the
governing equation in~\eqref{eq:detdyn} or~\eqref{eq:detndyn} and the
observable $f$: since the field $u$ is typically a complicated
function of $\vartheta$, establishing the conditions under which this
assumption holds will have to be done on a case-by-case basis. Note
that it guarantees that the set $\Omega(z)$ is simply connected with a
boundary that is $C^1$ for all $z\in \RR$, with inward pointing unit
normal at $\theta(z) \in \partial \Omega(z)$ given by
$\hat n (z)= \nabla F(\theta(z))/|\nabla F(\theta(z))|$. We could
relax the constraint $|\nabla F(\theta)|>0$, and allow e.g.~for the
sets $\Omega(z)$ to have several connected components (the number of
which could depend on $z$), but this requires to modify the argument
below.  Assumption~\ref{th:as3} allows us to introduce the tilted
measure
\begin{equation}
  \label{eq:21}
  d\mu_\eta(\theta) = \frac{e^{\<\eta, \theta\>} d\mu(\theta)
  }{\int_\Omega e^{\<\eta,  \theta\>} d\mu(\theta)} 
  = e^{\<\eta, \theta\>-S(\eta)} d\mu(\theta)\,.
\end{equation}
It is easy to see that the mean of $\mu_\eta$ is shifted compared to
that of $\mu$. A simple calculation shows that
\begin{equation}
  \label{eq:22}
  \int_\Omega \theta \, d\mu_\eta(\theta) =
  \nabla S(\eta)\,, 
\end{equation}
and this will allow us to pick $\eta$ such that the mean of $\mu_\eta$
is precisely at the point minimizing $I(\theta)$ in $\Omega(z)$. Note
that
\begin{equation}
  \label{eq:28}
  \Omega(z+\delta) \subseteq \Omega(z) \qquad \forall z\in\RR,\delta\ge0\,,
\end{equation}
and to establish~\eqref{eq:2} we will find conditions such that (i)
$\mu(\Omega(z))$ decreases fast with~$z$ and (ii) this probability is
dominated by a small region around a single point on
$\partial \Omega(z)$. This will require us to make additional
assumptions on the geometry of $\Omega(z)$ that we discuss next in
connection with properties of the rate function $I(\theta)$ defined
in~\eqref{eq:6bb}.

Letting
\begin{equation}
  \label{eq:2gb}
  \theta^\star(z) = \argmin_{\theta\in \Omega(z)} I(\theta)\,,
\end{equation}
we first make:
\begin{assumption}
  \label{th:as4}
  There exists a finite $z_0$ such that, $\forall z\ge z_0$,
  $\theta^\star:[z_0,\infty) \to \Omega$ is continuously
  differentiable and $I(\theta^\star(\cdot))$ is strictly increasing
  with $z$ with
    \begin{equation}
  \label{eq:32c}
  I(\theta^\star(z))\to \infty \quad \text{and} \quad |\nabla
  I(\theta^\star(z))|\ge K >0
  \quad \text{as}\quad z\to\infty.
    \end{equation}
\end{assumption}
\noindent
This assumption implies that $\theta^\star(z) \in \partial \Omega(z)$
for $z>z_0$, i.e.~we can replace~\eqref{eq:2gb} with
\begin{equation}
  \label{eq:2g}
  \theta^\star(z) = \argmin_{\theta\in \partial\Omega(z)} I(\theta)\,.
\end{equation}
The Euler Lagrange equation for~\eqref{eq:2g} is
\begin{equation}
  \label{eq:25}
  \nabla I(\theta^\star(z)) = \lambda \nabla F(\theta^\star(z))
\end{equation}
for some Lagrange multiplier $\lambda$.  Since by definition both $S$
and $I$ are convex functions, by the involution property of the
Legendre transform we have
\begin{equation}
  \label{eq:6inv}
  S(\eta) = \max_{\theta} \left( \<\eta, \theta\> - I(\theta)
  \right)\,,
\end{equation}
and this maximum is achieved at the solution of
\begin{equation}
  \label{eq:5}
  \eta = \nabla I(\theta)
\end{equation}
in $\theta$. Therefore if we define $\eta^\star(z)$ via
\begin{equation}
  \label{eq:7dual}
  \eta^\star(z) = \nabla I(\theta^\star(z)) 
\end{equation}
the mean of $\mu_{\eta^\star(z)}$ is
$\theta^\star(z)$. From~\eqref{eq:6inv} this also implies that
\begin{equation}
  \label{eq:27}
  \<\eta^\star(z), \theta^\star(z)\> - S(\eta^\star(z))  = I(\theta^\star(z))\,,
\end{equation}
which gives the following exact representation formula for $\mu(\Omega(z))$
\begin{equation}
  \label{eq:33}
  \begin{aligned}
    \mu(\Omega(z)) & = \int _{\Omega(z) } e^{S(\eta^\star(z))-
      \<\eta^\star(z),
      \theta\> } d\mu_{\eta^\star(z)}(\theta) \\
    & = e^{-I(\theta^\star(z))}\int _{\Omega(z) } e^{-
      \<\eta^\star(z), (\theta-\theta^\star(z))\> }
    d\mu_{\eta^\star(z)}(\theta)\,.
  \end{aligned}
\end{equation}
To proceed further we need to make some assumptions about
$\Omega(z)$. First:
\begin{assumption}
  \label{th:as5}
  For all $z\ge z_0$, the set $\Omega(z)$ is contained in the
  half-space whose boundary is tangent to $\Omega(z)$ at
  $\theta= \theta^\star(z)$, i.e.
\begin{equation}
  \label{eq:1h}
  \Omega(z) \subseteq \mathcal{H}(z) = \left\{ \theta \ : \< \hat
    n^\star(z), \theta- \theta^\star(z)\> \ge0\right\},
\end{equation}
where
$\hat n^\star (z)= \nabla F(\theta^\star(z))/|\nabla
F(\theta^\star(z))|$ denotes the inward pointing unit normal to
$\partial \Omega(z)$ at $\theta^\star(z)$.
\end{assumption}
\noindent
In the terminology of Ney\cite{Ney:1983iz}, it means that
$\theta^{\star}(z)$ is a dominating point in $\Omega(z)$.  If we
combine~\eqref{eq:25} and \eqref{eq:7dual} we deduce that
\begin{equation}
  \label{eq:30}
  \frac{\eta^\star (z)}{|\eta^\star(z)|} = 
  \frac{\nabla F(\theta^\star(z))}{|\nabla F(\theta^\star(z))|}
  = \hat n^\star(z)
\end{equation}
and as a result we can use Fubini's theorem to express \eqref{eq:33}
as
\begin{equation}
  \label{eq:4b}
  \begin{aligned}
    \mu(\Omega(z)) & = e^{-I(\theta^\star(z))}\int_0^\infty
    e^{-|\eta^\star(z)| s} |\eta^\star(z)| G(z,s)\,ds\,.
  \end{aligned}
\end{equation}
Here we defined
\begin{equation}
  \label{eq:10b}
  G(z,s)  =  \mu_{\eta^\star(z)}\left(\Omega(z)\setminus \mathcal{H} (z,s)\right),
\end{equation}
with
\begin{equation}
  \label{eq:9b}
  \mathcal{H} (z,s) = \left\{ \theta \ : \ \<\hat n^\star(z), (\theta - \theta^\star(z) -
    \hat n^\star(z) s)\> \ge 0\right\}\,.
\end{equation}
Note that in~\eqref{eq:4b} the lower limit of the integral is at $s=0$
by Assumption~\ref{th:as5}.
Since by definition we have
\begin{equation}
  \label{eq:6}
  \forall s>0 \ : \ G(z,s) \in (0,1), \qquad \forall s,s'>0, \ s'>s  \
  : \
  G(z,s')>G(z,s),
  \qquad \lim_{s\to0^+} G(z,s) = 0\,,
\end{equation}
from \eqref{eq:4b} we obtain the upper bound
\begin{equation}
  \label{eq:7}
  \mu(\Omega(z)) \le e^{-I(\theta^\star(z))}\int_0^\infty
  e^{-|\eta^\star(z)| s} |\eta^\star(z)| ds = e^{-I(\theta^\star(z))}\,,
\end{equation}
which implies
\begin{equation}
  \label{eq:10}
  \frac{\log \mu(\Omega(z))}{I(\theta^\star(z))} \le -1\,.
\end{equation}
To get a matching lower bound notice that for all $s_1>0$ we have
\begin{equation}
  \label{eq:9}
  \begin{aligned}
    \mu(\Omega(z)) & \ge e^{-I(\theta^\star(z))}\int_0^{s_1}
    e^{-|\eta^\star(z)| s} |\eta^\star(z)| G(z,s) ds\\
    & \ge e^{-I(\theta^\star(z))}
    G(z,s_1)\left(1-e^{-|\eta^\star(z)| s_1} \right)\\
    & \ge e^{-I(\theta^\star(z))}
    G(z,s_1)\frac{|\eta^\star(z)| s_1}{1+|\eta^\star(z)| s_1}\,.
  \end{aligned}
\end{equation}
Therefore if we make:
\begin{assumption}
  \label{th:as6} There exists $s_1>0$ such that
  \begin{equation}\label{eq:10ass}
    \lim_{z\to\infty} \frac{\log G(z, s_1) 
    }{I(\theta^\star(z))} = 0,
  \end{equation}
\end{assumption} 
\noindent 
for this $s_1$ we have (using also Assumption~\ref{th:as4} that
guarantees that $|\eta^{\star}(z)|\ge K>0$)
\begin{equation}
  \label{eq:9log}
  \begin{aligned}
    \frac{\log\mu(\Omega(z))}{I(\theta^\star(z))} & \ge -1
    +\frac{\log G(z, s_1) + \log \left( |\eta^\star(z)|
        s_1\right) -\log \left( 1+|\eta^\star(z)| s_1\right)
    }{I(\theta^\star(z))}\\
    & = -1 +\frac{\log G(z, s_1)
      -\log \left( 1+|\eta^\star(z)|^{-1} s^{-1}_1\right) }{I(\theta^\star(z))}\\
    & \to -1 \qquad \text{as \ \ $z\to\infty$}\,.
  \end{aligned}
\end{equation}
Combining~\eqref{eq:10} and~\eqref{eq:9log} we finally deduce
\begin{theorem}[Large deviation principle]\label{th:1}
  Under Assumptions~\ref{th:as1}--\ref{th:as6},  the following
  result holds:
\begin{equation}
  \label{eq:13b}
  \lim_{z\to\infty} \frac{\log P(z) }{I(\theta^\star(z))} =
  \lim_{z\to\infty} \frac{\log\mu(\Omega(z)) }{I(\theta^\star(z))} =
  -1.
\end{equation}
\end{theorem}
\noindent
Note that \eqref{eq:13b} is just a rephrasing of~\eqref{eq:2}.

It is useful to comment on the assumptions on $\Omega(z)$ that lead to
Theorem~\ref{th:1}. Assumption~\ref{th:as4} states that the event
$\{F(\vartheta) \ge z\}$ becomes rare as $z\to\infty$, which is
clearly necessary for an LDP to apply. Assumption~\ref{th:as5}
guarantees that all regions in $\Omega(z)$ remain much more unlikely
than $\theta^\star(z)$: this assumption can be relaxed, but at the
price of having to analyze more carefully how $I(\theta)$ behaves on
$\partial \Omega(z)$ and exclude that regions with lower likelihood
near this boundary accumulate and eventually dominate the probability.
Finally, Assumption~\ref{th:as6} is about the shape of the set
$\Omega(z)$ near $\theta^\star(z)$. Since the mean of
$\mu_{\eta^\star(z)}$ is $\theta^\star(z)$, we know that this measure
must have mass in a region around $\theta^\star(z) $ but we need to
make sure that this region has sufficient overlap with
$\Omega(z)$. For example, if for each $z\ge z_0$ we can insert in
$\Omega(z)$ a set that contains $\theta^\star(z)$ on its boundary and
is such that its volume remains finite as $z\to\infty$,
Assumption~\ref{th:as6} will automatically hold. On the other hand,
this assumption could fail for example if $\Omega(z)$ becomes
increasingly thin.  More discussion about this kind of geometric
assumptions can be found
e.g.~in~\cite{iltis1995sharp,kuelbs2000large}.

It is also interesting to note that~\eqref{eq:4b} offers a way to
derive asymptotic expansions for $\mu(\Omega(z))$ more refined
than~\eqref{eq:13b} if we assume that: (i) $|\eta^\star (z)|$ grows
with $z$, i.e.~we supplement~\eqref{eq:32c} with
\begin{equation}
  \label{eq:32d}
  |\eta^\star (z)| = |\nabla I(\theta^\star(z))| \to \infty
  \qquad \text{as \ $z\to\infty$\,;}
\end{equation} 
and (ii) $G(z,s)$ has a specific behavior near $s=0$ as
$z\to\infty$. For example, suppose that there is a $C>0$ such that
for all $u\ge0$
\begin{equation}
  \label{eq:12b}
  G(z,|\eta^\star(z)|^{-1} u)\sim C
  |\eta^\star(z)|^{-\alpha} u^\alpha 
  \qquad \text{with \  $\alpha>0$\ \ as\ \  $z, |\eta^\star(z)|\to\infty$}\,,
\end{equation}
where $f(z)\sim g(z)$ indicates that
$\lim_{z\to\infty} f(z)/g(z) = 1$. Then we have
\begin{equation}
  \label{eq:13c}
  \begin{aligned}
    P(z) = \mu(\Omega(z)) & = e^{-I(\theta^\star(z))}\int_0^\infty e^{-u}
    G\left(z, |\eta^\star(z)|^{-1} u\right) du\\
    &\sim
    e^{-I(\theta^\star(z))} C |\eta^\star(z)|^{-\alpha}\int_0^\infty e^{-u} 
    u^{\alpha} du\\
    & = C \Gamma(\alpha+1) |\eta^\star(z)|^{-\alpha} e^{-I(\theta^\star(z))}\,.
  \end{aligned}
\end{equation}
It is interesting to note that both \eqref{eq:4b} and~\eqref{eq:13c}
are consistent with $\vartheta|_{\Omega(z)}$ (outcome of the event
conditioned on $F(\vartheta)\ge z$) having fluctuations of order
$O(|\eta^\star(z)|^{-1})$ away from $\theta^\star(z)$ in the direction
parallel to $\eta^\star(z)$. Perpendicular to $\eta^\star(z)$ the
fluctuations remain of order $O(1)$ even as $z\to\infty$, but
integrating in these perpendicular directions only gives a
sub-exponential correction to $\mu(\Omega(z))$. This correction
depends on the geometry of the hypersurface $\partial\Omega(z)$ (in
particular on its curvature) near $\theta^\star(z)$. This is what is
accounted for in~\eqref{eq:13c}, and this picture will be confirmed in
the numerical examples below.

\subsubsection*{Illustration: Gaussian measure with linear observable}
Let us illustrate the LDT optimization in the simple case of a
Gaussian random variable $\vartheta$ with mean 0 and covariance
$\text{Id}$, taking values $\theta \in \RR^N$. If we consider a linear
observable
\begin{equation}
  \label{eq:g3}
  \mF(\theta) = \< b, \theta\>, \qquad b \in \RR^N\,,
\end{equation}
we have
\begin{equation}
  \label{eq:g4}
  \begin{aligned}
    \PP(\<b, \vartheta\> \ge z ) & = (2\pi)^{-N/2} \int_{\<b, \theta\> \ge z}
    \exp\left(-\tfrac12 |\theta|^2\right) d\theta\,,
  \end{aligned}
\end{equation}
and a direct calculation shows that
\begin{equation}\label{eq:g4b}
  \PP(\<b,\theta\>\ge z) = \tfrac12\erfc\left( \frac{z}{\sqrt2 |b|}
  \right) \sim  (2\pi)^{-1/2} |b|z^{-1}
  \exp\left(-\tfrac12 |b|^{-2} z^2\right) \quad \text{as} \ \ z\to\infty\,.
\end{equation}
Let us check that the LDP derived above is consistent with this
result.  Here
\begin{equation}
  \label{eq:11}
  I(\theta) = \tfrac12 |\theta|^2, \qquad S(\eta) = \tfrac12|\eta|^2.
\end{equation}
If we minimize $I(\theta)$ subject to $\<\theta,b\>\ge z$, we deduce
\begin{equation}
  \label{eq:12}
  \theta^\star (z) = z |b|^{-2}  b\qquad \text{and}\qquad  I(\theta^\star(z))
  = \tfrac12 |b|^{-2} z^2.
\end{equation}
Comparing this result with~\eqref{eq:g4b} we see that it is consistent
with the prediction in~\eqref{eq:13b}.

We can also test what the theory can say beyond the log-asymptotic
estimate. Here, the planar condition corresponding to
$\Omega(z) = \mathcal{H}(z)$ is exactly fulfilled by linearity of
$F(\theta)= \< b,\theta\>$.  We need to estimate $G(z,|\eta^\star(z) |^{-1})$ as
$z\to\infty$. From \eqref{eq:7dual} and \eqref{eq:27} we have that
\begin{equation}
  \label{eq:g9a}
  \eta^\star(z) = \nabla I(\theta^\star(z))=\theta^\star(z) = z
  |b|^{-2} b, \qquad S(\eta^\star(z)) = \tfrac12 z^2 |b|^{-2} \,,
\end{equation}
and the tilted measure~\eqref{eq:21} at $\eta=\eta^\star(z)$ reads
\begin{equation}
  \label{eq:14}
  d\mu_{\eta^\star(z)}(\theta) = (2\pi)^{-N/2} \exp\left( - \tfrac12|\theta|^2
    + z b|b|^{-2} \<b,\theta\>
    - \tfrac12 z^2|b|^{-2} \right) d\theta\,.
\end{equation}
Using \eqref{eq:g9a}, we obtain
\begin{equation}\label{eq:g9b}
	\begin{aligned}
          G(z,s) & = \int_{z\le \<b,\theta\>\le
            z+s} d\mu_{\eta^\star(z)}(\theta)\\
          & =(2\pi)^{-1/2}  \int_0^{s} \exp\left( - \tfrac12 u^2 \right) du\\
          & = \tfrac12\erf\left(\tfrac12\sqrt{2} s\right)\,.
	\end{aligned}
\end{equation}
As a result
\begin{equation}
  \label{eq:13}
  G(z,|\eta^{\star}(z)|^{-1}s)  
  \sim (2\pi)^{-1/2} |\eta^{\star}(z)|^{-1}s =  (2\pi)^{-1/2} |b| s
  z^{-1}
  \qquad \text{as} \ \ z\to\infty.
\end{equation}
Comparing with~\eqref{eq:12b}, we see that here $C=(2\pi)^{-1/2}$ and
$\alpha=1$.  Therefore~\eqref{eq:13c} agrees with~\eqref{eq:g4b} as
expected.

\section{Numerical aspects}
\label{sec:optimization}

Here we review how to numerically perform the minimization
in~\eqref{eq:2} and thereby estimate $P(z)$ -- the method can be
straightforwardly generalized to consider also the minimization
associated with the calculation of \eqref{eq:3}
or~\eqref{eq:expectationsndyn}. We impose the constraint
$f(u(T))\ge z$ by adding a Lagrange multiplier term to \eqref{eq:2},
so that the minimization can be rephrased in Hamiltonian formalism by
\cite{troltzsch2010optimal,borzi2011computational}:
\begin{equation}\label{eq:Hamilton-a}
   E(u,\theta) = I(\theta)-\lambda f(u(T)))\,,
\end{equation}
where $u(T)$ should itself be viewed as a function of $\theta$
obtained by solving~\eqref{eq:detdyn} with $\vartheta=\theta$, that is
\begin{equation}
  \label{eq:detdyn00}
  \partial_t u = b(u,\theta), \qquad u(t=0) = u_0(\theta).
\end{equation}
The minimization of \eqref{eq:Hamilton-a} with $u(T)$ obtained
from~\eqref{eq:detdyn00} can be performed via steepest descent with
adaptive step (line search). This requires to compute the gradient of
$E$ with respect to $\theta$, which can be achieved in two ways: by
the direct and the adjoint
methods~\cite{borzi2011computational,plessix2006review}. These steps are described
next.

\subsection{Gradient Calculation}
\label{sec:gradcalc}

\subsubsection{Direct method}
\label{sec:Jacobian}

The gradient of the cost function with respect to the control reads:
\begin{equation}
  \label{eq:gradE}
  \nabla_\theta E(u(T,\theta),\theta) = \partial_\theta E + \left(\partial_\theta
    u(T,\theta)\right)^T \partial_u E = \nabla_\theta I -
  \lambda \, J^T(T,\theta) \, \partial_u f(u(T,\theta))\,,
\end{equation}
where $J = \partial_ \theta u$ is the Jacobian---componentwise
$J_{i,j}= \partial u_i/\partial\theta_j$. An evolution equation for
$J$ can be obtained by differentiating~\eqref{eq:detdyn00} with
respect to $\theta$:
\begin{equation}
  \label{eq:jacobian}
  \partial_t J = \partial_u b \, J + \partial_\theta b, \qquad	
  J(0) = \nabla_\theta \,u_0.
\end{equation}
Summing up, given the current state of the control, $\theta^n$, we
calculate the gradient of the objective function via:
\begin{enumerate}
\item \textit{Field estimation:} Obtain the current field $u^n$ by solving
  \begin{equation}\label{eq:forward}
    \partial_t u^n = b(u^n,\theta^n), \qquad u^n(0) = u_0(\theta^n)\,.
  \end{equation}
\item \textit{Jacobian estimation:} Obtain the Jacobian $J^n$ by solving
\begin{equation}
  \label{eq:Jacobian2}
  \partial_t J^n = \partial_u b(u^n,\theta^n) \, J^n
  + \partial_\theta b(u^n,\theta^n), \qquad	
  J^n(0) = \nabla_\theta \,u_0(\theta^n).
\end{equation}
\item \textit{Gradient calculation:} Compute the 
  gradient $(\nabla_\theta E)^n$ via
  \begin{equation}
    (\nabla_\theta E)^n = \nabla_\theta I(\theta^n) - \lambda \, (J^n(T))^T \, \partial_u
    f(u^n(T)).
  \end{equation}
\end{enumerate}

\subsubsection{Adjoint method}
\label{sec:adjoint}

Let us introduce the adjoint field $\mu(t)$ solution of
\begin{equation}
  \label{eq:adjoint}
  \partial_t \mu = - (\partial_u b)^T\mu, 
  \qquad \mu(T,\theta) = \lambda \partial_u f(u(T,\theta)).
\end{equation}
Using this equation as well as the transpose of~\eqref{eq:jacobian} we deduce
\begin{equation}
  \label{eq:15}
  \begin{aligned}
    \partial_t (J^T \mu) & = \partial_t J^T \mu + J^T \partial_t \mu
    \\
    &=  J^T (\partial _u b)^T \mu  + (\partial_\theta b)^T \mu - J^T (\partial_u
    b)^T\mu = (\partial_\theta b)^T \mu\,.
  \end{aligned}
\end{equation}
As a result
\begin{equation}
  \label{eq:17}
  \int_0^T (\partial_\theta b)^T \mu dt = J^T(T) \mu(T) - J^T(0)
  \mu(0) = \lambda  J^T(T,\theta) \partial_u f(u(T,\theta)) - (\nabla
  _\theta u_0)^T \mu(0,\theta).
\end{equation}
This expression offers a way to write the gradient of the objective
function in~\eqref{eq:gradE} as
\begin{equation}
  \label{eq:gradE3}
  \nabla_\theta E = \nabla_\theta I - (\nabla
  _\theta u_0)^T  \mu(0,\theta) -\int_0^T (\partial_\theta b)^T \mu \, dt\,.
\end{equation}
Using this expression instead of~\eqref{eq:gradE} is computationally
advantageous because it avoid the calculation of the Jacobian~$J$ --
note in particular that the adjoint field $\mu$ has the same
dimensions as $u$, independent of the dimensions of the space
$\Omega$. The price to pay is the field $ u$ must be computed and
stored separately since~\eqref{eq:adjoint} for $\mu$ must be solved
backward in time. Summarizing, the gradient of the objective function
is now calculated via:
\begin{enumerate}
\item \textit{Field estimation:} Obtain the current field $u^n$ by solving
  \begin{equation}\label{eq:forwardbis}
    \partial_t u^n = b(u^n,\theta^n), \qquad u^n(0) = u_0(\theta^n).\,
  \end{equation}
\item \textit{Adjoint field estimation:} Obtain the adjoint field $\mu^n$ by solving
  \begin{equation}\label{eq:backward}
    \partial_t \mu^n = - (\partial_u b(u^n,\theta^n))^T\mu^n, \qquad 
    \mu^n(T) = \lambda \partial_u f(u^n(T)).
  \end{equation}
\item \textit{Gradient calculation:} Compute the gradient 
  $(\nabla_\theta E)^n$ via
  \begin{equation}\label{eq:gradE4}
    (\nabla_\theta E)^n = \nabla_\theta I(\theta^n) - (\nabla
    _\theta u_0(\theta^n))^T \mu^n(0) -\int_0^T (\partial_\theta
    b(u^n(t),\theta^n))^T \mu^n(t) \, dt \,.
  \end{equation}
\end{enumerate}
Note that equations \eqref{eq:forwardbis} for $u$ and
\eqref{eq:backward} for $\mu$ are adjoint in both space and time. As a
result the numerical simulation of these equations has to be done with
care, as the integration scheme used for one equation needs to be the
adjoint of the other. This is preferably done by using schemes that
are self-adjoint. For recent literature on the topic we refer the
reader to~\cite{wilcox2015discretely, hager2000runge,
  walther2007automatic}.

\subsection{Descent with pre-conditioning of the gradient}
\label{sec:descent}

Once we have calculated the gradient of the objective function at
$\theta^n$, we can make a downhill step in the cost function
landscape using 
\begin{itemize}
\item[(4)] \textit{Descent step with pre-conditioning:}
  \begin{equation}
    \label{eq:update}
    \theta^{n+1} = \theta^n - \alpha^n B^n (\nabla_\theta E)^n \,,
  \end{equation}
\end{itemize}
where $B^n$ is a pre-conditioning $M\times M$ matrix (recall
that $\theta \in \Omega \subseteq \RR^M$), and $\alpha^n>0$ is the
step size that is tuned optimally at each iteration via line search:
this can be done using classical merit functions as discussed
in~\cite{wright1999numerical}.

The estimate of the matrix $B^n$ deserves some further
comments. Ideally, $B^n$ should be the inverse of the Hessian of the
objective function $E(\theta^n)$, but this Hessian is typically
difficult to calculate. Therefore, a simpler solution is to use the
Hessian of the prior $I(\theta^n)$, which in the case of a Gaussian
measure is simply the inverse covariance matrix $C^{-1}$ (which is
independent of $\theta$. Since this estimate coincides with the
Hessian of $E(\theta^n)$ only when $\lambda=0$, it will deteriorate
when $\lambda$ increases and the pre-conditioning may become
inefficient. If that is the case, it may be useful to switch to
``quasi-Newton'' methods such as the BFGS algorithm, or the
Limited-Memory BFGS algorithm when $M$ is very large ($>100$). In the
applications treated in this paper, the naive pre-conditioning
depending only on the prior $I(\theta)$ turned out to be sufficient to
perform the optimization efficiently.

Since we are typically interested in calculating~\eqref{eq:2} for a
range of values of~$z$, instead of fixing~$z$ and trying to determine
the corresponding Lagrange multiplier~$\lambda$ in~\eqref{eq:gradE},
it is easier to vary~$\lambda$ and determine \textit{a~posteriori}
which value of~$z$ this leads to. Indeed this offers a parametric
representation of $\theta^\star(z)$ via
\begin{equation}
  \label{eq:4}
  \theta^\star(z(\lambda)) = \tilde \theta^\star(\lambda),
  \qquad z(\lambda)  = f(u(T,\tilde\theta^\star(\lambda))\,,
\end{equation}
where $\tilde \theta^\star(\lambda)$ is the minimizer of $E(\theta)$
at $\lambda$ fixed. We can then also calculate
$I(\theta^\star(z(\lambda))) = I(\tilde \theta^\star(\lambda))$ and
estimate $P(z(\lambda))\asymp \exp(-I(\tilde \theta^\star(\lambda)))$.

\section{Applications}
\label{sec:applications}

\subsection{Elasticity of an heterogeneous rod}
\label{sec:rod}

In this section we study a model for a one-dimensional rod with random
elasticity coefficient subject to a prescribed external mechanical
forcing (i.e. pulling at one end). Even though this model (or
generalizations thereof) may be of interest in actual applications
(e.g.~as a coarse-grained model of DNA
stretching~\cite{bustamante1994entropic,cluzel1996dna,lankavs2000sequence}),
it is primarily used here as a simple illustrative example of the
tools and concepts introduced in Secs.~\ref{sec:LDTmethod}
and~\ref{sec:optimization}. In particular, we use LDT to locate the
most likely configurations leading to extreme responses and we show
that such realizations dominate the statistics asymptotically.

In the case of forcing increasing linearly in time, we are able to
derive analytical results which are used to validate our numerical
method. We also study the extreme events that occur under a nonlinear
forcing, when no analytical solution is available.

\subsubsection{Continuous model with random structure}
Consider a one-dimensional elastic rod of length 1 that is being
pulled at one end with a time-dependent force and whose energy is
specified in terms of its displacement field $u:[0,1] \to \RR$ via
\begin{equation}\label{eq:cont-Ham-rod}
  V(u,t) = \frac12 \int_0^1 \mD(x)\,|\partial_x u|^2\,dx - r(t)u(1)\,,
\end{equation}
where the first term is the total internal energy of the rod and the
second term is the external energy (negative of the work potential);
$\mD(x)>0$ is the elasticity coefficient, assumed to be spatially
dependent, and $r(t)$ is a prescribed external forcing protocol acting
on the right end of the rod -- the specific form of $r(t)$ will be
introduced later. The dynamics of the rod is governed by the
Euler-Lagrange equation associated with~\eqref{eq:cont-Ham-rod}:
\begin{equation}\label{eq:cont-rod}
  \partial^2_t u = \partial_x(\mD(x)\partial_x u)\quad x \in (0,1)\,,
\end{equation}
with initial conditions to be prescribed later and boundary conditions
\begin{equation}\label{eq:bound-rod}
  u(t,0)=0\,,\quad \mD(1)\partial_x u(t,1)=r(t)\,, \quad \forall t\ge0\,.
\end{equation}
In order to introduce uncertainty in the model we make the elasticity
random, i.e.~we take $\mD(x) \equiv \mD(x,\vartheta)$. Here we will
assume that $\mD(x,\vartheta)$ is piecewise constant over blocks of
size $1/M$ for some $M\in \NN$, with independent values in each
block. Specifically, we take:
\begin{equation}
  \label{eq:coarse}
  \mD(x,\vartheta) = \sum_{k=1}^M \varphi_k(x) g(\vartheta_k )\,,
\end{equation}
where  the functions $\{\varphi_k\}_{k=1}^M$ are given by
\begin{equation}\label{eq:basis}
  \varphi_k(x)= \left\{ 
    \begin{aligned}
      &1 \qquad \text{if }\ \  M^{-1}(k-1) \le x< M^{-1} k \\ 
      &0\qquad \text{otherwise}
    \end{aligned}\right.;
\end{equation}
$g$ is a given function; and  $\{\vartheta_k\}_{k=1}^M $
are \text{i.i.d.} random variables. Below we will consider two cases:

\smallskip

\paragraph{\textit{Case 1.}} Here we assume that $g:(0,\infty) \to (0,\infty)$ with
\begin{equation}
  \label{eq:37}
  g(u) = u^{-1}
\end{equation}
and we take the variable $\{\vartheta_k\}_{k=1}^M$ to be exponentially
distributed, i.e.
\begin{equation}
  \label{eq:8}
  \PP(\vartheta_k \ge \theta_k) = e^{-\alpha \theta_k}, \qquad \theta_k \ge
  0, \quad \alpha >0\,.
\end{equation}
This choice implies that
\begin{equation}
  \label{eq:16}
  S(\eta) = \log \EE e^{\<\eta,\vartheta\>} = -\sum_{k=1}^M \log
  (1-\alpha^{-1}\eta_k), \qquad \eta_k < \alpha \quad \forall k=1,\ldots, M,
\end{equation}
so that
\begin{equation}
  \label{eq:18}
  I(\theta) = \sum_{k=1}^M \left(\alpha \theta_k - 1 - \log \theta_k
  \right) \qquad \theta_k >0 \quad \forall k=1,\ldots, M\,.
\end{equation}

\smallskip

\paragraph{\textit{Case 2.}} Here we assume that $g:\RR\to(0,\infty)$
with
\begin{equation}\label{eq:PDFchain}
  g(u) = \tfrac12u +
  \sqrt{\tfrac14u^2+1}, 
\end{equation}
and we take the variable $\{\vartheta_k\}_{k=1}^M$ to be normally
distributed with variance $\sigma^2>0$, i.e.
\begin{equation}
  \label{eq:38}
  \vartheta_k = \mathcal{N} (0,\sigma^2) 
\end{equation}
This choice implies that
\begin{equation}
  \label{eq:39}
  S(\eta)  = \tfrac12 \sum_{k=1}^N \sigma^2 \eta_k^2, \qquad
  I(\theta) = \tfrac12 \sum_{k=1}^N \sigma^{-2} \theta_k^2\,.
\end{equation}

Given this random input, our aim is to investigate the statistics of
the displacement of the right end of the rod at time $T$: this amount
to considering the observable $f(u(T))=u(T,1)$, and studying the
behavior of
\begin{equation}
  \label{eq:29}
  P(z) = \PP(u(T,1, \vartheta)\ge z) \qquad \text{for} \ \ z \gg 1.
\end{equation}
Below we will analyze the behavior of this quantity in two cases, when
the forcing $r(t)$ in~\eqref{eq:cont-Ham-rod} is linear in $t$ and
when it is not -- the first situation is amenable to analytical
treatment whereas the second is not in general. Note that in both
situations, the behavior of $P(z)$ for large $z$ will depend on how
fast $g(u)$ decays to zero: due to the shape of $g$ this will
depend on the right tail of the distribution of $\vartheta_k$ in Case
1 and on its left tail in Case 2.

\subsubsection{Discrete model}
\label{sec:discrete}

To perform the numerics, we need to consider a spatially discretized
version of the model above. We do so by introducing the discrete
energy
\begin{equation}\label{eq:discr-Ham-rod}
  V(u,t) = \frac12
  \sum_{j=0}^{N-1} \mD_{j+1}(\vartheta) \frac{(u_{j+1}-u_j)^2}{\Delta x} - r(t)\, u_N \,,
\end{equation}
in which $u_j = u(j\Delta x)$, $\mD_j= \mD(j\Delta x)$,
$\Delta x =1/N$.  Alternatively, \eqref{eq:discr-Ham-rod} can be
thought of as the energy for a system of $N+1$ beads $u_j$ connected
by $N$ springs with random spring constants $\mD_j(\vartheta)$.  The
dynamics obeys the system of ODEs
\begin{equation}\label{eq:rod}
  \partial_t^2{u_j} =	 \frac{\mD_{j+1}}{\Delta
    x^2}(u_{j+1}-u_j) 
  - \frac{\mD_j}{\Delta x^2}(u_j - u_{j-1})\,,\qquad j=1,...,N-1\,,
\end{equation}
with fixed boundary condition $u_0=0$ at the left end and dynamic
boundary condition
\begin{equation}\label{eq:dynb1}
  \partial_t^2u_N = - \frac{\mD_N}{\Delta x^2}(u_N - u_{N-1}) + \frac{r(t)}{\Delta x} 
\end{equation}
at the right end. 
We will pick $N=P M$ for some $P\in \NN$, so that by our choice for
$\mD(x,\vartheta)$ in~\eqref{eq:coarse} we have
\begin{equation}
  \label{eq:40}
  \mD_j(\vartheta) = g(\vartheta_k) \qquad \text{for} \ \ \lceil j/P
    \rceil= k, \ \ j=1,\dots,N, \ \ k = 1\ldots,M.
\end{equation}

Since we focus on the statistics of the observable
$f(u(T))=u_N(T)=u(T,1)$ that measures the displacement at time $T$ of
the right end point with respect to its initial position, the cost
function is
\begin{equation}\label{eq:Ediscr}
  E(u,\theta) = I(\theta) - \lambda\, u_N(T) \,,
\end{equation}
to optimize on the parameters $\{\theta_k\}_{k=1}^M$. We will
minimize~\eqref{eq:Ediscr} using the adjoint method to compute the
gradient. As shown in the Appendix, the adjoint equations read
\begin{equation}\label{eq:adj-discr}
  \partial_t^2 \mu_j = \frac{\mD_{j+1}}{\Delta x^2}(\mu_{j+1}-\mu_j) 
  - \frac{\mD_j}{\Delta x^2}(\mu_j - \mu_{j-1})\,,\qquad j=1,...,N-1\,,
\end{equation}
with conditions at the boundaries given by
\begin{equation}\label{eq:BCadj}
  \mu_0(t) = 0 \,, \quad \partial_t^2\mu_N 
  = -\frac{\mD_N}{\Delta x^2} (\mu_N-\mu_{N-1})\,,
\end{equation}
and final conditions
\begin{equation}\label{eq:ICadj}
	\mu_j(T) = 0, \quad \partial_t \mu_j(T) = \lambda \delta_{j,N}\,.
\end{equation}
The gradient of the cost function can be expressed as
\begin{equation}\label{eq:gradRod1}
  \nabla_\theta E(u(\theta),\theta) = \nabla I(\theta) - G^T\,\nabla \mD(\theta)\,,
\end{equation}
where $\nabla \mD(\theta)$ is the $N\times M$ tensor with entries
$\partial \mD_j(\theta)/\partial \theta_k$, $j=1,\ldots, N$,
$k=1,\ldots, M$, and $G$ is a vector with entries
\begin{equation}\label{eq:gradRod2}
  G_j = \int_0^T \frac{u_{j}-u_{j-1}}{\Delta x}
  \frac{\mu_{j}-\mu_{j-1}}{\Delta x}\,dt\,, \qquad j=1,\ldots, N\,.
\end{equation}

\subsubsection{Linear forcing}
Assume that $r(t) = a t $ for some $a>0$ and as initial conditions
for~\eqref{eq:in-rod} take
\begin{equation}\label{eq:in-rod1}
  u(0,x)=0\,,\quad \partial_t u(0,x)=
  a \int_0^x \frac{dx'}{\mathcal D(x',\theta)} \,, \quad \forall x \in [0,1]\,.
\end{equation}
The solution to~\eqref{eq:cont-rod} equipped with the boundary
conditions in~\eqref{eq:bound-rod} is
\begin{equation}
  \label{eq:19}
  u(t,x,\vartheta) = a t \int_0^x \frac{dx'}{\mD(x',\vartheta)}\,.
\end{equation}
Let us consider the implications of this formula in {\it Case 1},
which is suitable to derive analytical results.  Eq.~\eqref{eq:19}
implies that
\begin{equation}
  \label{eq:20}
  u(T,1, \vartheta) = aT \int_0^1\frac{dx'}{\mD(x',\vartheta)} = 
  \frac{aT}{M} \sum_{k=1}^M \vartheta_k\,,
\end{equation}
where we used the specific form of $\mD(x,\vartheta)$ given
in~\eqref{eq:coarse} with $g$ given in~\eqref{eq:37}. Note that since
the discrete equivalent to the initial conditions~\eqref{eq:IC} is
\begin{equation}\label{eq:IC1}
  u_j(0) = 0, \qquad \partial_t u_j(0) = \frac{a}{M}\sum_{k=1}^j \theta_k\,,
\end{equation}
the result~\eqref{eq:20} also holds for the discretized model, i.e.~we
have
\begin{equation}
  \label{eq:20b}
  u_N(T, \vartheta) =
  \frac{aT}{M} \sum_{k=1}^M \vartheta_k\,.
\end{equation}
From~\eqref{eq:8}, this implies that
$u(T,1, \vartheta)=u_N(T,\vartheta)$ follows a gamma distribution with
shape parameter~$M$ and rate parameter $\alpha M (aT)^{-1}$:
\begin{equation}
  \label{eq:23}
  \begin{aligned}
    P(z) &= \int_z^\infty \frac{(\alpha M (aT)^{-1})^M
      u^{M-1}}{(M-1)!}
    e^{-\alpha M (aT)^{-1} u} du\\
    & = \frac{1}{(M-1)!}\Gamma\left(M,\alpha M(aT)^{-1}z\right)\,,
  \end{aligned}
\end{equation}
where $\Gamma(\cdot,\cdot)$ is the upper incomplete Gamma function.
When $z\gg1$ with $M$ fixed, \eqref{eq:23} gives
\begin{equation}
  \label{eq:24}
  P(z) \sim \frac{(\alpha M (aT)^{-1}   z)^{M-1}}{(M-1)!} e^{-\alpha M (aT)
    ^{-1}  z}\,,
\end{equation}
meaning that
\begin{equation}
  \label{eq:31}
  \log P(z) \sim -\alpha M (aT)^{-1} z +(M-1) \log (\alpha M (aT)^{-1}   z)
  - \log(M-1)!.
\end{equation}
In this last expression the second and third terms at the right hand
side are subdominant over the first, $\alpha M (aT) ^{-1} z$, and
disappear in the limit as $z\to\infty$. It is useful to keep this
terms for comparison with the result~\eqref{eq:13b} in
Theorem~\ref{th:1} and the result~\eqref{eq:13c}, which we do next.

If we solve
\begin{equation}
  \label{eq:26}
  \min I(\theta) = \min \sum_{k=1}^M \left(\alpha \theta_k - 1 - \log \alpha\theta_k
  \right) \qquad \text{subject to} \qquad u(T,1, \theta) = \frac{aT}{M}
  \sum_{k=1}^M \theta_k = z\,,
\end{equation}
we get
\begin{equation}
  \label{eq:32}
  \theta^\star_k(z) = (aT)^{-1} z \qquad \text{for} \ \ k=1,\ldots, M\,.
\end{equation}
As a result
\begin{equation}
  \label{eq:34}
  I(\theta^\star(z)) = M\left(\alpha  (aT)^{-1} z -1 -
  \log(\alpha (aT)^{-1} z)\right)\,,
\end{equation}
which from~\eqref{eq:31} is consistent with
$\log P(z) \sim - I(\theta^\star(z)) $ as $z\to\infty$, as predicted
by~\eqref{eq:13b}. Note also that here
\begin{equation}
  \label{eq:35}
  \eta^\star _k(z) = \partial_{\theta_k} I(\theta^\star(z)) = \alpha -
   aT z^{-1} \qquad \text{for} \ \ k=1,\ldots, M\,.
\end{equation}
Since this implies that $|\eta^\star _k(z)|\to \alpha $ as
$z\to\infty$, this means that the condition in~\eqref{eq:32d} is not
satisfied here.
\begin{figure}
\centering
  \centering
  \includegraphics[width=.5\linewidth]{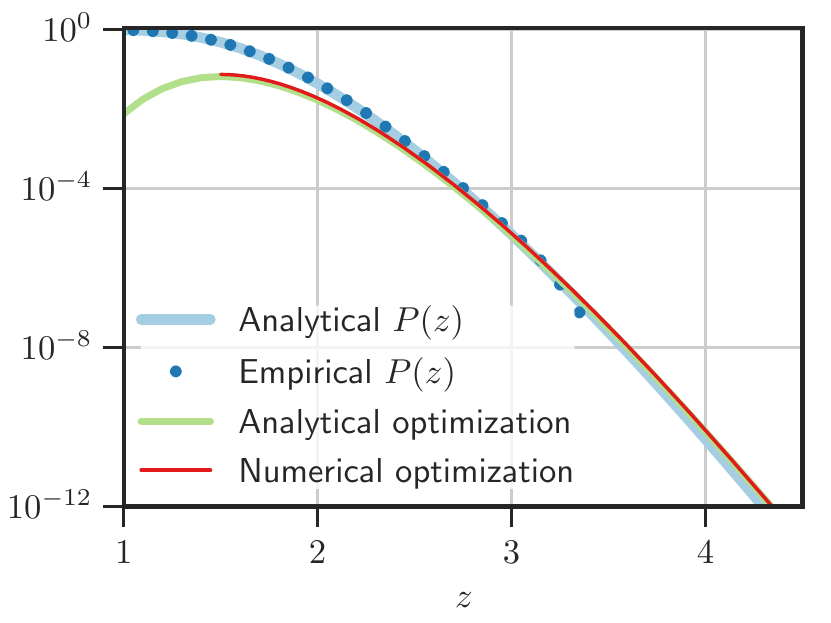}%
  \includegraphics[width=.5\linewidth]{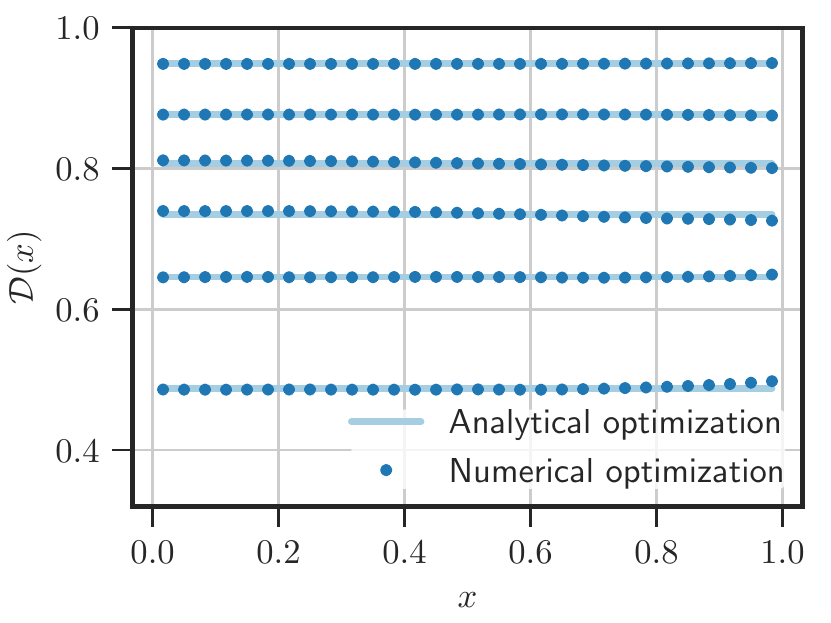}
  \caption{Linear forcing with $a=0.1$, final time $T=15$, initial
    conditions~\eqref{eq:IC1}, and the statistical prior of {\it
      Case~1}. The numerics are performed with $M=N=30$. Left panel:
    Comparison between the exact expression for $P(z)$
    in~\eqref{eq:23}, the empirical MC estimate with $2\times10^7$
    samples, the analytical LDT estimate~\eqref{eq:34}, and the LDT
    estimate obtained via numerical optimization. Right panel:
    Comparison between the analytical~\eqref{eq:32} and the numerical
    instantons, for $z=1.58,1.71,1.85,2.04,2.32,3.08$ from top to
    bottom.}
\label{fig:1}
\end{figure}

In Fig.~\ref{fig:1} we compare the asymptotic estimate~\eqref{eq:34}
with the exact expression~\eqref{eq:23}. We also  check that the
numerical optimization is consistent with the analytical one, which is
important to validate the numerical code described below.

\subsubsection{Nonlinear forcing}
\begin{figure}
  \centering
{\includegraphics[width=.6\linewidth]{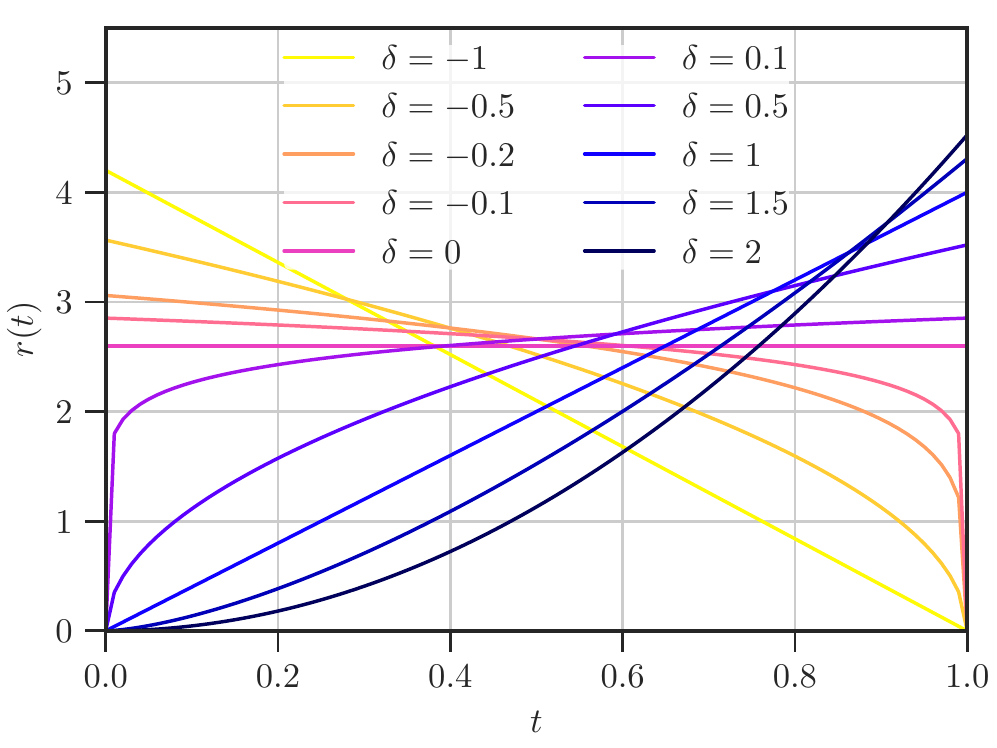}%
}
\caption{The forcing protocols $r_\delta(t)$ in~\eqref{eq:41}, which
  are decreasing functions of $t$ when $\delta<0$ and increasing
  functions when $\delta>0$.}
\label{fig:rod1}
\end{figure}

\begin{figure}
\centering
 { \includegraphics[width=.6\linewidth]{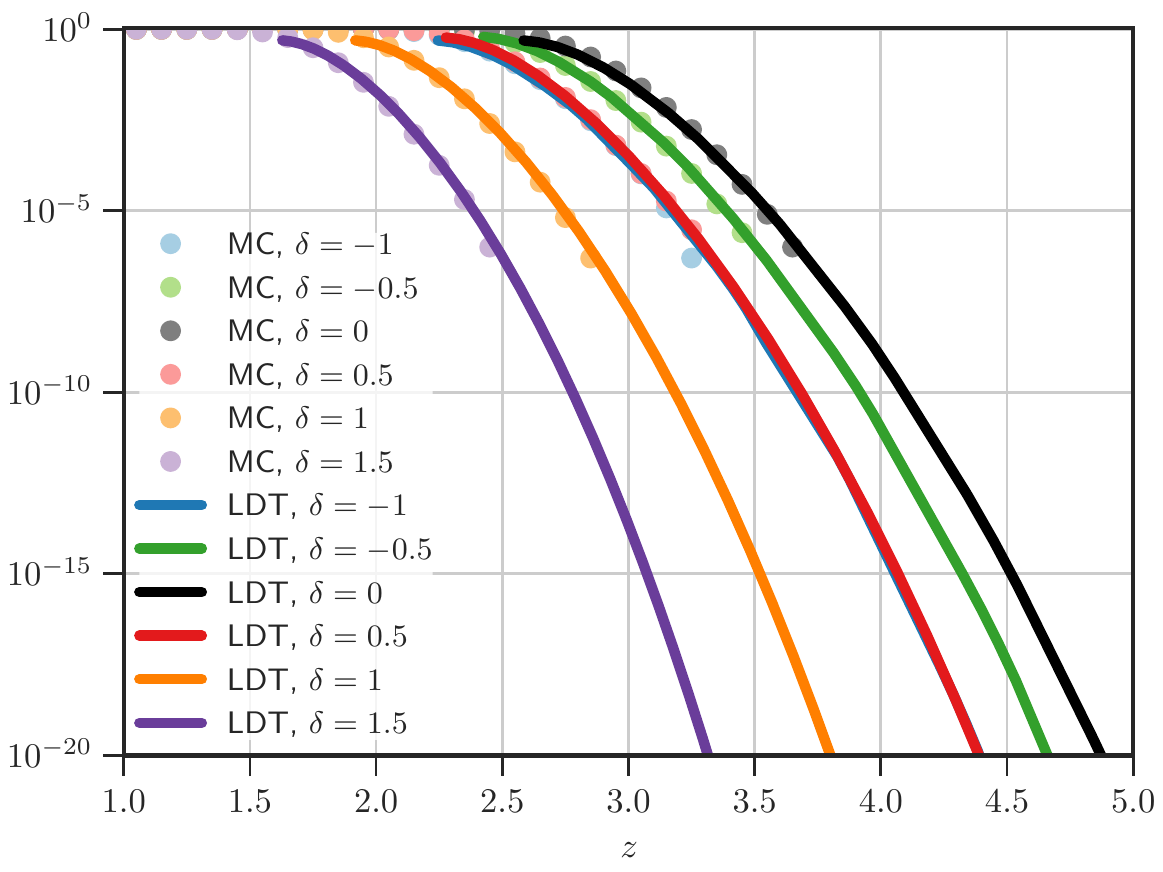}}
 \caption{Comparison between the empirical distributions $P(z)$
   obtained via MC sampling and their LDT estimate. The sampling works
   down to events whose probability is about the inverse of the MC
   sampling size, while the LDT optimization allows us to extend the
   tails to much smaller probabilities.}
\label{fig:2}
\end{figure}

Next we consider nonlinear forcing protocols of the type
\begin{equation}
  r(t)=at^\beta \qquad \text{and}\qquad  r(t)=a(T-t)^\beta \qquad
  \text{both with} \quad a, \beta>0\,.\label{eq:41}\
\end{equation}
Letting $s=+1$ if $r(t) = at^\beta $ and $s=-1$ if
$r(t)=a(T-t)^\beta$, we will use $r_\delta(t)$ with $\delta=s\beta$ as
shorthand to describe the family of forcing protocols. They are show
in Fig.~\ref{fig:rod1}.

As initial conditions for \eqref{eq:cont-rod} we take
\begin{equation}\label{eq:in-rod}
	u(0,x)=0\,,\quad \partial_t u(0,x)=0\,, \quad \forall x \in [0,1]\,.
\end{equation}
At discrete level these initial
conditions read
\begin{equation}\label{eq:IC}
  u_j(0) = 0, \qquad \partial_t u_j(0) = 0\,.
\end{equation}
In this section we restrict ourselves to {\it Case~2} and we use
$M=N=30$ and final time $T=1$.  Observing that the mean elasticity
$\EE \left(\mD(x)\right)=1$ (as for {\it Case 1}), the average
velocity of propagation of the waves along the bar is also $1$. Thus,
$1$ is the average time that a signal takes to propagate from the
right end to the left end. This means that taking $T=1$ we are
considering a short transient strongly out of equilibrium, where the
random structure will contribute in a non-homogeneous way.

To integrate~\eqref{eq:rod} and~\eqref{eq:dynb1} numerically, we use a
velocity-Verlet integrator, which is of second order, symplectic, and
time reversible, with a time step of $10^{-3}$. The optimization is
performed as described in Sec.~\ref{sec:optimization},
using~\eqref{eq:gradRod1} and~\eqref{eq:gradRod2}.

\begin{figure}
  \centering
  \includegraphics[width=320pt]{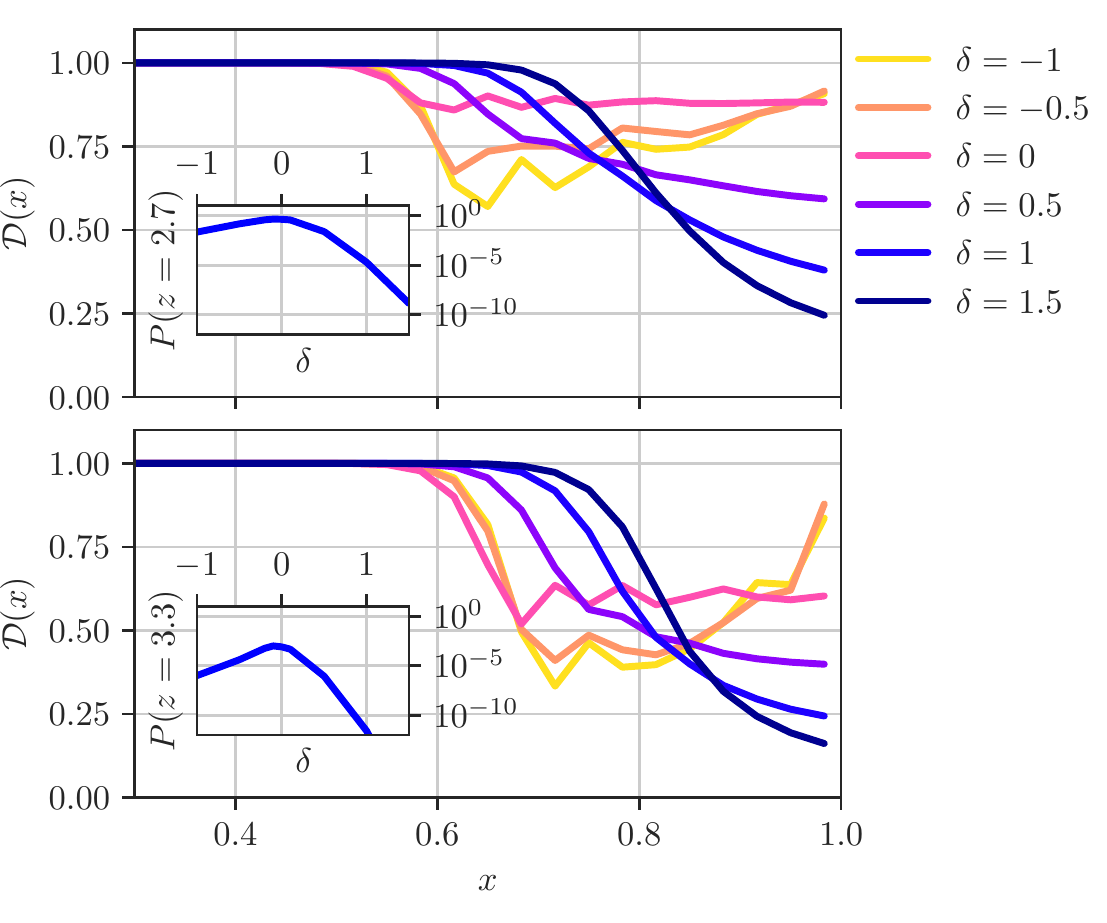}
  \caption{Top panel: Elasticity structure of the instantons for
    $z\ge2.7$, for the different protocols labeled
    by~$\delta$. Inset: the probability $P(z=2.7)$ as a function of
    the forcing protocol. Bottom panel: Same as in the top panel, but
    for $z\ge3.3$. Inset: the probability $P(z=3.3)$ as a function of
    $\delta$ in the forcing protocol.}
\label{fig:rod5}
\end{figure}

Let us now describe our results.  In Fig.~\ref{fig:2} the LDT
estimates of $P(z)$ are compared to the empirical estimates obtained
via MC with $2\times10^6$ samples, showing good agreement.  Next we
look at the specific elasticity structure of the optimizers,
$\mathcal{D}(x,\theta^\star(z))$. These are shown in
Fig.~\ref{fig:rod5}. As can be seen, the region that is relevant for
having an extreme extension $u(T=1,1)$ occupies only the right half of
the space domain, independent of the protocol. This makes sense since
on average the signal takes a time $1$ to cross the whole domain: For
a point $x_0$ to influence $u(T=1,1)$ the signal needs to have time to
propagate to $x=1$. As a result, the points on the left side will not
have the possibility to influence the dynamics at all, and the optimal
state of $\mathcal{D}(x,\theta)$ is determined by mere minimization of
$I(\theta)$ with no dynamical constraint. In contrast, on the right
side of the domain, $\mathcal{D}(x,\theta)$ must take low values to
allow for large values of $u(T=1,1)$ -- since these low values are
unlikely, this also account for the drop in probability observed in
Fig.~\ref{fig:2}. Fig.~\ref{fig:rod5} also indicates that
$\mathcal{D}(x,\theta^\star(z))$ depends on the forcing protocol. This
dependency can again be interpreted intuitively by realizing that the
region that impacts $u(T,1)$ the most will be the one that is reached
by a strong signal (i.e.~the propagation front of the most intense
part of the forcing) and is able to send a strong feedback back to the
right end at final time -- this feedback is what is accounted for by
the backward evolution of the adjoint equation in the
optimization. So, the earlier the most intense part of the forcing
takes place, the further from the right end a low elasticity peak
appears. This explains why going towards negative $\delta$ the
low-elasticity peak moves to the left in Fig.~\ref{fig:rod5}, and the
constant forcing ($\delta=0$) is the one where the low elasticity
contribution is the most uniformly distributed.

Note that in this framework it is possible to compare how likely
the protocols are to produce extreme realizations of a given size, as
shown in the insets in Fig.~\ref{fig:rod5}. In this sense, the
constant protocol appears to be the optimal one. This is consistent
with the fact that $\delta=0$ is the highest curve in
Fig.~\ref{fig:2}.

\begin{figure}
  \centering
 \includegraphics[width=.8\linewidth]{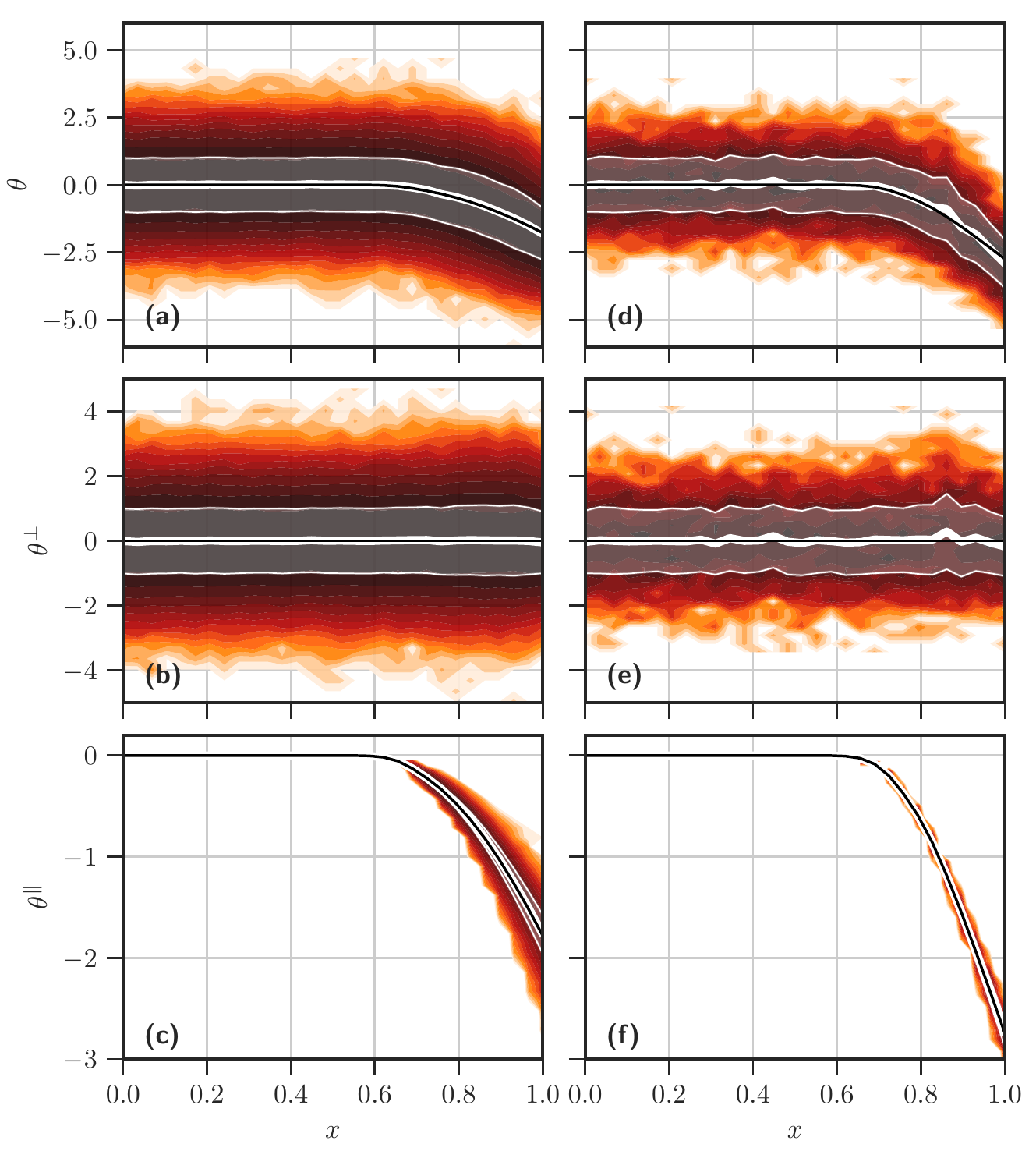}
 \caption{Comparison between the instanton $\theta^\star(z)$ (black
   solid line) and the Monte Carlo sampling on the distribution of
   $\vartheta$, conditioned on $u_N(1)=u(1,1)\ge z$ (color map with
   intensity proportional to the empirical probability density; thick
   white line = mean; thin white lines = 1 standard deviation range around the
   mean). Left panels: $z=2.10$, right panels: $z=2.40$. The top
   panels show the full data: the instanton agrees with the mean, but
   the variance does not substantially change going to more extreme
   events.  The two central panels show the fluctuations perpendicular
   to $\eta^\star(z)$, confirming that their amplitude is independent
   of the size of the event (left and right panels have the same
   variance) and homogeneous in space. The bottom panels show the
   fluctuations in the direction parallel to $\eta^\star(z)$,
   indicating that their amplitude decreases as $z$ increases, as
   predicted by the theory in Sec.~\ref{sec:LDTmethod}.}
\label{fig:3}
\end{figure}
\begin{figure}
\centering
\includegraphics[width=.6\linewidth]{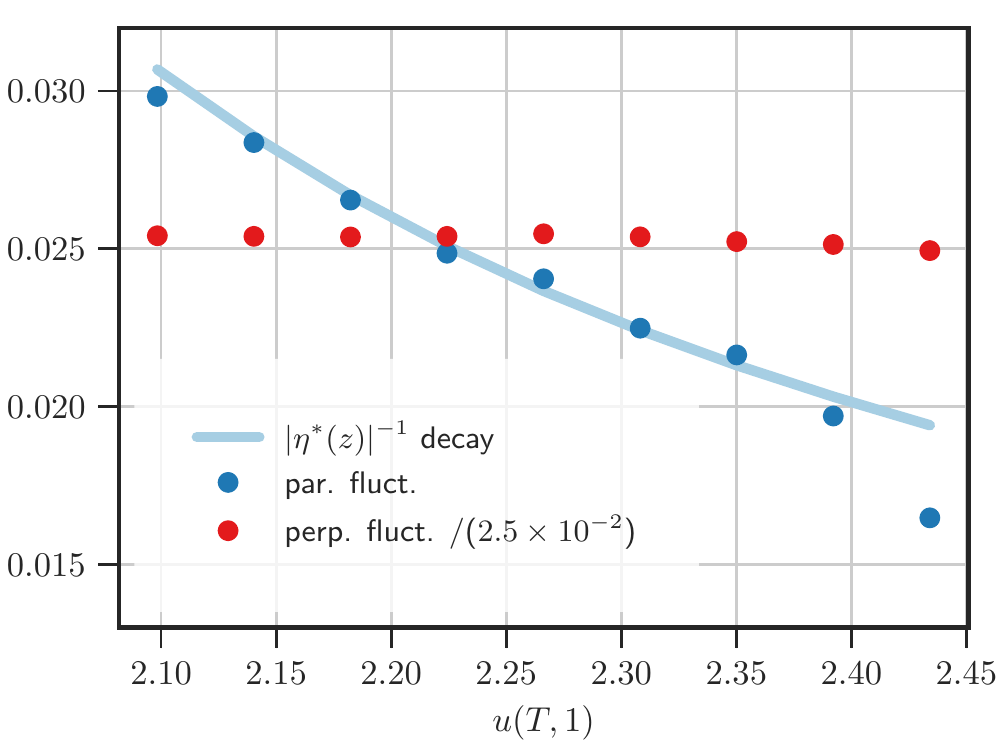}
\caption{Increasing $z$, the fluctuations in the direction
  perpendicular to $\eta^\star(z)$ stay constant, whereas in the
  parallel direction they scale as $O(|\eta^\star(z)|^{-1})$. Both
  behaviors are predicted analytically and here confirmed
  numerically.}
\label{fig:4}
\end{figure}

To further clarify the role of the instantons and why they dominate
the dynamics and the statistics of the extreme events, it is useful to
``filter'' the conditional events such that $u(T,1)\ge z$ in the
following way: First, we fix a size $z$ and generate via MC a large
set of $\vartheta$ such that $u(T,1,\vartheta)\ge z$. Second, we
average over such conditional set to obtain the mean conditional event
and its fluctuations around the mean, which is generally very close to
the instanton $\theta^\star(z)$. Third, we decompose the fluctuations
$\vartheta - \theta^\star(z)$ into the components parallel and
perpendicular to $\eta^\star(z)$, i.e.~the normal to the hypersurface
$\Omega(z)$. This procedure is then repeated for various~$z$.

In Fig.~\ref{fig:3} we show the outcome of this analysis for the
protocol with $\delta=1.5$ and for two different values of $z$ --
analogous results hold for the other kinds of forcing as well. As can
be seen the average event $u(T,1)\ge z$ lies on top of the instanton
$\theta^\star(z)$, with fluctuations independent of the size of the
event and also of the position along the rod (upper panels). The
decomposition shows that the components perpendicular to
$\eta^\star(z)$ are independent of the size of the event, and
basically independent of the dynamics too. Their mean and standard
deviation are the mean and the standard deviation of the unconstrained
random variables $\vartheta$ (central panels). In contrast, the
parallel fluctuations are small and tend to zero as $z$ increases
(bottom panels).  The scaling of the fluctuations is analyzed in more
detail in Fig.~\ref{fig:4}, which shows that they are $O(1)$ in the
direction perpendicular to $\eta^\star(z)$ and
$O(|\eta^\star(z)|^{-1})$ in the direction parallel to it, consistent
with the theoretical predictions.

\subsection{Extreme events in optical turbulence}\label{sec:NLS}
\subsubsection{The 1D NLSE and the LDT formalism}

The nonlinear Schr\"odinger equation (NLSE) in one dimension arises in
a variety of different contexts such as surface gravity
waves~\cite{zakharov:1968,onorato-residori-bortolozzo-etal:2013},
nonlinear fiber optics~\cite{akhmediev2013recent},
plasmas~\cite{bailung2011observation} and Bose-Einstein
condensates~\cite{gross:1961,pitaevsky:1961}.  Here we will focus on
applications of NLSE in nonlinear optics, a domain that has seen
exciting experimental developments in recent
years~\cite{kibler2010peregrine, suret2016single,
  tikan2017universality}.  Specifically, we study the problem of the
onset of rogue waves out of a bath of random waves taken as initial
condition for NLSE, which is a key question in \textit{integrable
  turbulence}~\cite{zakharov2009turbulence,randoux2014intermittency,%
  agafontsev2015integrable,cousins2016reduced,farazmand-sapsis:2017}.

In non-dimensional units, the 1D NLSE for the envelope of a light beam
propagating in an optical fiber reads
\begin{equation}
  \label{eq:NLS}
    \partial_{\xi} \Psi = i \frac{1}{2}\Psi_{\tau\tau} + i
    |\Psi|^2\Psi\,,
    \quad \tau \in \Gamma,
\end{equation}
where \mbox{$\Gamma=[0,T]$}, with periodic boundary conditions
$\Psi(\xi,0)=\Psi(\xi,T)$, and a suitable initial condition
$\Psi(0,\tau)=\Psi_0(\tau)$, at the input end of the fiber $\xi=0$.
The non-dimensional distance~$\xi$, time~$\tau$, and envelope~$\Psi$
are related to the respective physical quantities $x$, $t$, and $\psi$
via characteristic constants that depend on the specifics of the
optical fiber: $x = \mathcal L_0 \xi$, $t = \mathcal T_0 \tau$ and
$\psi = \sqrt{\mathcal P_0}\Psi$. For instance, if we pick
$\mathcal T_0 = 5\text{ ps}$, $\mathcal L_0 = 0.5\text{ km}$,
$\mathcal P_0 = 0.5\text{ mW}$, the NLSE~\eqref{eq:NLS} models an
optical fiber with dispersion
$|\beta_2|=\mathcal T_0^2/\mathcal L_0 = 50\text{ ps}^2 \text{km}^{-1}$
and nonlinearity $\gamma=1/(\mathcal L_0\mathcal P_0)=4 \text{ km}^{-1}\text{mW}^{-1}$. 

Let us denote by $\{\hat \Psi_n\}_{n\in\ZZ}$ the Fourier component of
$\{\Psi(\tau)\}_{\tau\in[0,T]}$, i.e.
\begin{equation}
  \label{eq:Fourier}
  \hat \Psi_n=\frac1T \int_0^T e^{-i\omega_n \tau} \Psi(\tau)
  d\tau,\qquad
  \Psi(\tau) =  \sum_{n=-\infty}^{+\infty} e^{i\omega_n \tau} \hat \Psi_n\,,
\end{equation}
where $\omega_n=2\pi n/T$ and $n\in\mathbb{Z}$.  Equation
\eqref{eq:NLS} is derived under the \textit{quasi-monochromatic}
assumption, meaning that the spectrum $\hat C_n$ defined as
\begin{equation}
  \hat C_n=\frac1T \int_0^T e^{-i\omega_n \tau} C(\tau)
  d\tau\,,\qquad C(\tau-\tau')=\EE(\Psi_0(\tau)\bar \Psi_0(\tau'))\,,
\end{equation}
must be narrow -- here and below the bar denoting complex
conjugation. We will consider a Gaussian spectrum with 
\begin{equation}
  \label{eq:spectrum}
  \hat C_n = \mathcal{A} e^{-\omega_n^2/(2\Delta)}\,\qquad \mathcal{A}
    >0, \qquad \Delta >0,  \qquad -M \le n \le M, \quad M>0,
\end{equation}
and $\hat C_n=0$ for $|n|> M$.  Assuming that the initial
$\Psi(0,\tau)$ is a Gaussian field with mean zero and covariance
$C(\tau-\tau') $, this implies the representation
\begin{equation}
  \label{eq:ICFourier}
  \Psi(0,\tau,\vartheta) = \sum_{n=-M}^M  e^{i\omega_n \tau}
  \hat{C}_n^{1/2}\vartheta_n,
\end{equation}
where $\vartheta_n$ are complex Gaussian variables with mean zero and
covariance $\EE\vartheta_n\bar\vartheta_m=\delta_{m,n}$,
$\EE \vartheta_n \vartheta_m = \EE\bar \vartheta_n \bar \vartheta_m =
0\,.$ Note that the spectral amplitude is related to the optical power
$P(\xi,\tau)=|\psi(\xi,\tau)|^2$ (statistically homogeneous in $\tau$)
via $\mathcal{A}=\EE(P)/\sum_n e^{-\omega_n^2/(2\Delta)}$. The initial
statistical state of the system is thus completely determined given
the two parameters $\Delta$ and $\EE(P)$, and the average power
$\EE(P)$ is relevant to optical experiments -- it also enjoys the
property of being invariant under the NLSE evolution in the variable
$\xi$, i.e.~it can be measured at the input or at the output of the
optical fiber, equivalently.

In the set-up above, we will investigate extreme fluctuations of the
optical power at the output of the optical fiber ($\xi=L$). Recalling
that $|\Psi(L,\tau)|=\sqrt{P(L,\tau)}$, this amounts to looking at the
statistics of
\begin{equation}
  \label{eq:36}
  f(\Psi(\vartheta))= \max_{\tau\in\Gamma}|\Psi(L,\tau,\vartheta)| \,,
  \qquad L >0.
\end{equation}
Analyzing this observable using the framework developed in
Secs.~\ref{sec:LDTmethod} and \ref{sec:optimization} amounts to
minimizing  the cost function (this is~\eqref{eq:Hamilton-a}) 
\begin{equation}
  \label{eq:cost-waves}
  E(\Psi,\theta) = I(\theta) - \lambda f(\Psi) \qquad \text{with}
  \qquad I(\theta)=\tfrac12 \sum_{n=-M}^M \left|\theta_n\right|^2.
\end{equation}
This minimization must be performed on the $2\times(2M+1)$-dimensional
space $\Omega\subseteq\mathbb{C}^{2M+1}$ of the initial conditions.
The gradient of the cost function~\eqref{eq:cost-waves} is given by
\begin{equation}
  \label{eq:gradE-waves}
  \nabla_\theta E(\Psi(\theta),\theta) = \nabla_\theta I(\theta)
  + \Re (J(L,\tau_*))^T \frac{\Re (\Psi(L,\tau_*))}{|\Psi(L,\tau_*)|}
  + \Im (J(L,\tau_*))^T \frac{\Im (\Psi(L,\tau_*))}{|\Psi(L,\tau_*)|}\,,
\end{equation}
where $ \Psi(L,\tau_*)\equiv\max_{\tau\in\Gamma}
|\Psi(L,\tau)|\,$. The field $\Psi$ is evolved with~\eqref{eq:NLS} and
the initial condition depends on the point $\theta\in\Omega$ through
the mapping $\Psi(0,\theta)$ defined in~\eqref{eq:ICFourier}, with the
difference that here $\theta$ is no longer random. The matrix $J$
(also complex) evolves according to
\begin{equation}
  \label{eq:Jmnls}
  \partial_{\xi} J(\xi,\tau) = \int_0^L d\xi' \,
  \bigg(\frac{\delta b(\Psi(\xi))}{\delta \Psi(\xi')} \, J(\xi',\tau) + \frac{\delta b(\Psi(\xi))}{\delta \bar \Psi(\xi')} \, \bar J(\xi',\tau)\bigg) \,,
\end{equation}
where $b(\Psi(\xi)) )$ is a shorthand for the right hand side
of~\eqref{eq:NLS}: explicitly
\begin{equation}
  \label{eq:bmnls}
  \int_0^L d\xi' \,\frac{\delta b(\xi)}{\delta \Psi(\xi')}\, J(\xi')
  = \bigg(\frac{i}{2}\partial_{\tau\tau} + 2 i  |\Psi(\xi)|^2\bigg) J(\xi)\,,
\end{equation}
\begin{equation}
 \int_0^L d\xi' \,\frac{\delta b(\xi)}{\delta \bar \Psi (\xi')} \,
 \bar J(\xi')
 =  i \big(\Psi(\xi)\big)^2 J(\xi)\,.
\end{equation}
The initial condition for \eqref{eq:Jmnls} is
\begin{equation}
	\label{eq:incondJ}
	J(\xi=0,\theta) = \nabla_\theta \Psi(0,\theta)\,.
\end{equation}
Before turning to the results, let us explain how the numerical
simulations were performed. Equations~\eqref{eq:NLS}
and~\eqref{eq:Jmnls} were evolved from $\xi=0$ to $\xi=L$ (up to
$L=0.2$) using the pseudo-spectral second order Runge-Kutta
exponential-time-differencing method
(ETDRK2)~\cite{cox:2002,kassam:2005} with step $d\xi=5\times10^{-4}$
on a periodic box $[0,T]$ discretized by $2^{12}$ equidistant grid
points. The size $T=30$ is found large enough for the boundary
conditions to not affect the statistics on the spatio-temporal scales
considered. Each Monte Carlo simulation involves $10^6$ realizations
of the random initial data constructed via~\eqref{eq:ICFourier}, with
$M=45$. Adding more modes to the initial condition does not affect the
results in any significant way. The minimization was performed in the
space $\Omega$ (with high dimension $2\times (2M+1) = 182$). This step
was carried out via steepest descent with adaptive step (line search)
and preconditioning of the gradient, using the covariance of the
initial condition as metric, as explained in
Sec.~\ref{sec:optimization}.

\subsubsection{Results}

For generality, we present the results for the normalized field
$A(\xi,\tau)=\Psi(\xi,\tau)/\sqrt{\EE(P)}$ using non-dimensional
units. One can easily obtain the physical dimensions by applying the
straightforward transformations given below equation
\eqref{eq:NLS}. Four sets of parameters have been chosen to explore
different regimes: In Set 1, we take $\Delta=\pi$, $\EE (P) = 5/4$; in
Set 2, $\Delta=\pi/2$, $\EE (P) = 5/4$; in Set 3, $\Delta=3\pi/2$,
$\EE (P) = 5/4$; and in Set 4, $\Delta=\pi$, $\EE (P) = 5/9$.

\begin{figure}
\centering
  \includegraphics[width=.8\linewidth]{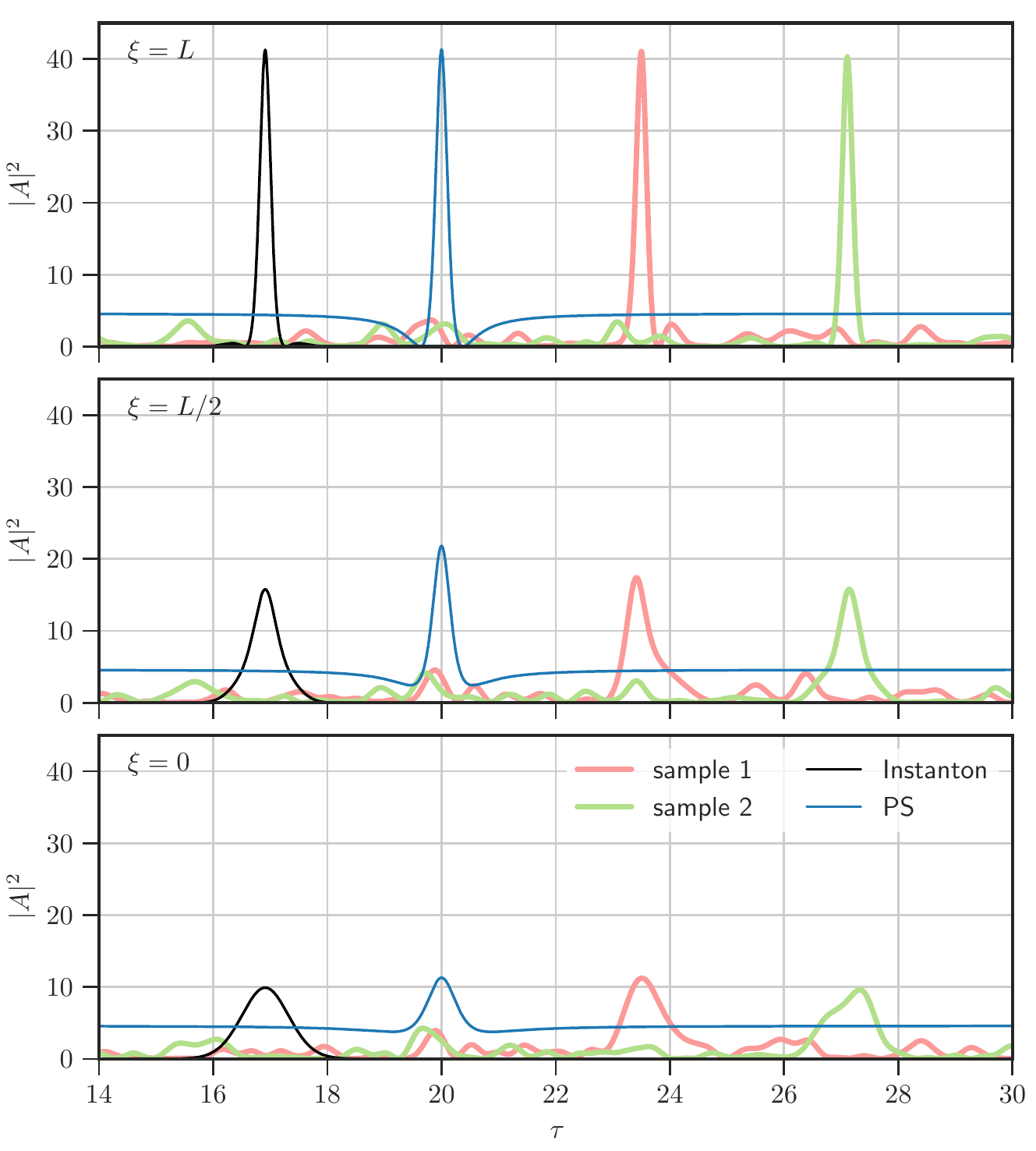}
  \caption{Set~1: The paths of occurrence of two extreme events plotted
    are compared with the instanton and the Peregrine solution reaching
    the same maximum power at $\xi=L$. Shown is the quantity
    $|A(\xi,\tau)|^2$, i.e.~the power in units of average power, at
    three different locations ($L=0.2$). The solution are shifted away
    from one another for clarity, exploiting homogeneity in~$\tau$.}
\label{fig:6}
\end{figure}

There have been recent claims, supported by both numerical and
experimental evidence~\cite{suret2016single, tikan2017universality},
about the universality of the Peregrine Soliton (PS) as a pathway to
optical rogue waves out of a random background. For this reason, we
carried out a comparison between the instantons and the PS. In
Fig.~\ref{fig:6}, the path of occurrence of two extreme events is shown for
Set~1, selected among the events in the random sampling with maximum
power amplification $|A|^2=P/\EE(P)$ exceeding a value of $40$. The
instanton and the PS reaching the same power amplification are also
plotted.
\begin{figure}
\centering
  \includegraphics[width=.6\linewidth]{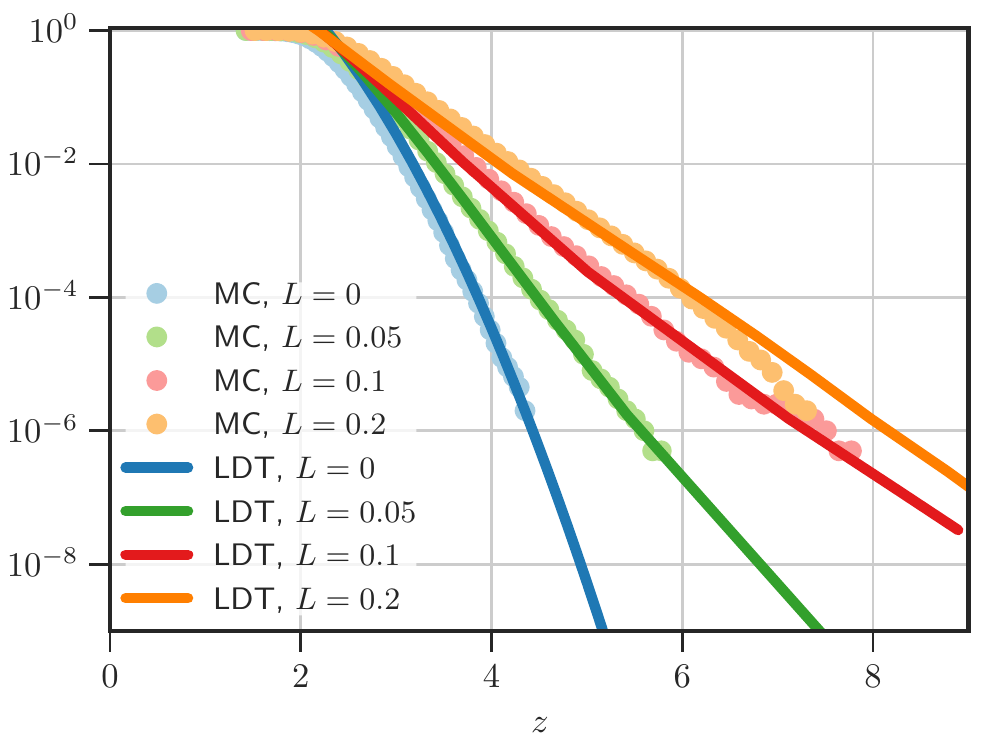}
  \caption{Set~1: Comparison between the probability distributions of
    $\max_\tau |A(L,\tau)|$ in the periodic time window $[0,T]$
    obtained by MC with $10^6$ samples, and their corresponding LDT
    estimates computed using the optimization method. The plot
    captures the tail fattening due to the NLSE dynamics, as the
    output point is taken at increasing distance $L$ from the
    input. The rogue-wave threshold is $|A|_{RW}\simeq 2.8$\,. The
    characteristic length of emergence of the coherent structures is
    $L_c=0.2$, compatible with the observed tail fattening.}
\label{fig:7}
\end{figure}
\begin{figure}
  \centering
  \includegraphics[width=.8\linewidth]{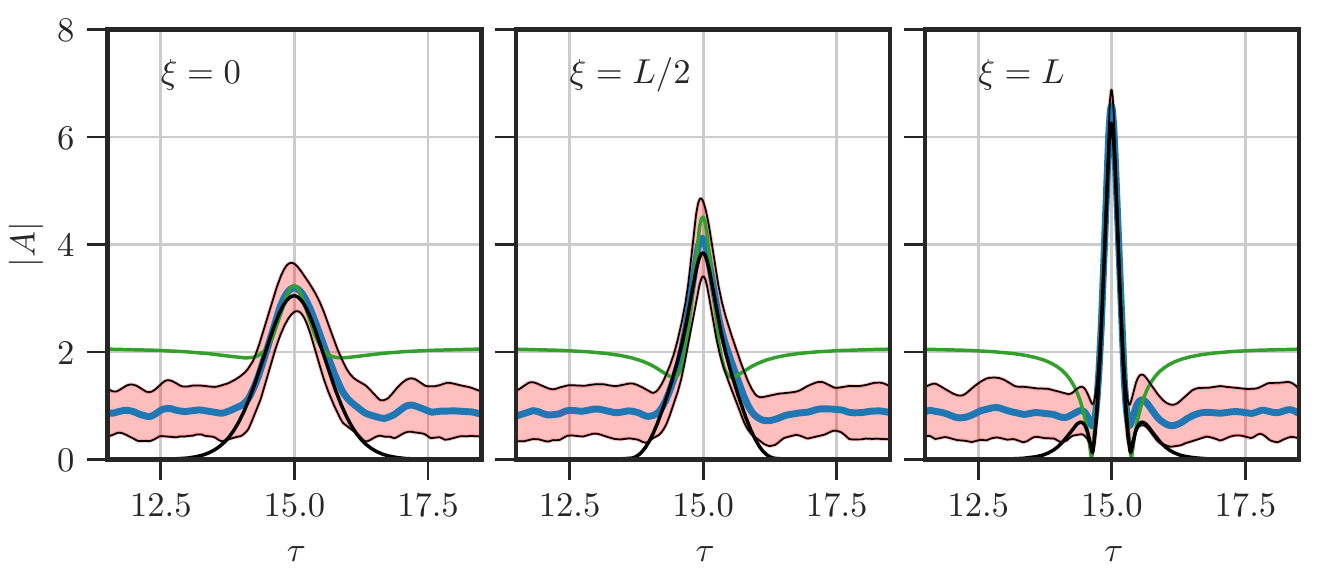}
  \caption{Set~1: Results of the conditioning on the sampling for
    $\max_{\tau\in\Gamma} |A(L,\tau)|\ge z=6.25$, with $L=0.2$.  Shown
    is the average of the conditional event (blue line), surrounded by
    the 1 std range (red area). The instanton (black line) is the
    optimal event reaching maximal intensity $A=z$ at the output point
    $\xi=L$. The PS is also represented (green line), normalized to
    have intensity $z$ at the point of maximal space-time
    focusing. From left to right, the panels are at $\xi=0$, $\xi=L/2$
    and $\xi=L$.}
\label{fig:8}
\end{figure}

In Fig.~\ref{fig:7} the probability
$P(z)=\PP(\max_\tau|A(L,\tau)|\ge z)$ is shown for various values of
$L$, showing good agreement between the results from MC sampling and
those from LDT optimization.  A rough estimate for the onset threshold
of optical rogue waves is
$|A|_{RW}=4\sqrt{2/\pi}\EE(|A|)\simeq 2.8$~\cite{el2018spontaneous},
independently of the set considered because of the use of the
normalized variable $A$.  As can be seen, the focusing NLSE increases
the probability of large excursions of $|A(L,\tau)|$ compared to its
initial Gaussian value with expectation
$\EE(|A(L=0,\tau)|) = \sqrt{\pi/4}$. This happens gradually as the
distance $L$ separating the input from the output increases. The tail
fattening can be interpreted quantitatively in terms of the typical
lengths of the coherent structures of NLSE. Defining the linear length
as $L_{\text{lin}}=2/\Delta^2$ and the nonlinear length as
$L_{\text{nlin}}=1/\EE (P)$, the typical length of emergence of a
coherent structure starting from a small hump is
$L_c = \tfrac12 \sqrt{L_{\text{lin}}L_{\text{nlin}}}$. This gives
$L_c=0.2$ for Set 1, in good agreement with the width of the spatial
transient over which the fast tail fattening takes place.

The asymptotic agreement of the probabilities shown in
Fig.~\ref{fig:7} is a numerical evidence that the focusing NLSE
\eqref{eq:NLS} with random initial data \eqref{eq:ICFourier} satisfies
an LDP. Additional support for the LDP is found in Fig.~\ref{fig:8},
where we compare the instanton with the sampling mean. Looking at the
signal to noise ratio, one sees that the events reaching a certain
extreme amplification are all very similar. According to the results
in Sec.~\ref{sec:LDTmethod}, these events are expected to have typical
fluctuations in the direction perpendicular to the instanton in the
space $\Omega$: notice how away from the focusing region (determined
by the direction perpendicular to the instanton because there the
instanton is vanishing) the observable $|A|$ fluctuates with standard
deviation $\sqrt{\EE(P)/2}\sqrt{(4-\pi)/2}/\sqrt{\EE(P)}\simeq0.57$
around the expected value
$\sqrt{\pi/2}\sqrt{\EE(P)/2}/\sqrt{\EE(P)}\simeq 0.89$, exactly as
expected for typical events. Instead, the extreme size of the event is
due to the component parallel to the instanton in $\Omega$, with small
fluctuations in this direction: As a matter of fact, in the focusing
region (determined by the component parallel to the instanton) the
signal to noise ratio becomes very big, meaning that, as $z$
increases, the extreme rogue waves with $\max_\tau|A(\tau,L)|\ge z$
become closer to the instanton reaching $\max_\tau|A(\tau,L)|= z$.

\begin{figure}
\centering
  \includegraphics[width=.8\linewidth]{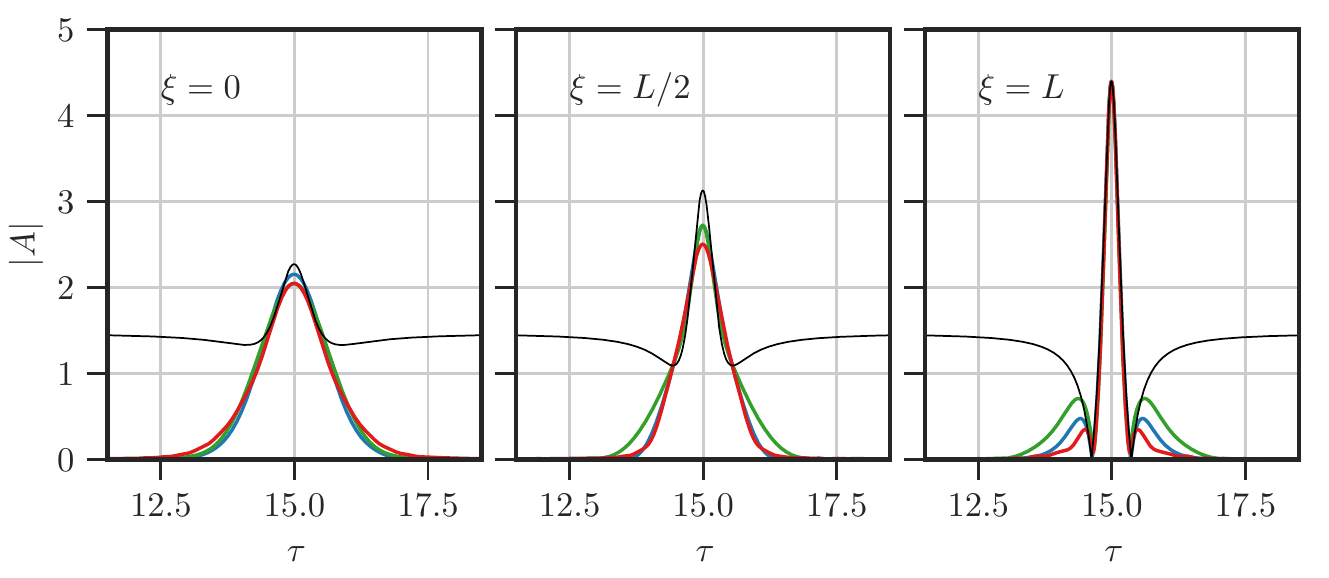}
  \caption{Snapshots at increasing spatial coordinate from left to
    right ($\xi=0$, $\xi=0.1$, $\xi=0.2$) of instantons reaching the
    same peak intensity, for the three sets of parameters with
    different spectral width: Set 1 ($\Delta=\pi$) in red; Set 2
    ($\Delta=\pi/2$) in green; and Set 3 ($\Delta=3\pi/2$) in
    blue. The PS reaching the same final height (at the point of
    maximal focusing) is also plotted in black. For all the profiles,
    striking agreement is observed around the point of maximal
    focusing in space-time, while significant differences are observed
    away from that point.}
\label{fig:10}
\end{figure}

Interestingly, from the knowledge of the LDT tails for a particular
configuration of the parameters $\Delta$ and $\EE(P)$ we can derive
the LDT tails for any combination of $\Delta$ and $\EE(P)$, using only
analytical transformations. This is possible thanks to two properties:
First, the scale invariance of the NLSE; second, the way the parameter
$\EE(P)$ appears in the cost function \eqref{eq:cost-waves}. Indeed the term
$I(\theta)$ is independent of $\EE(P)$, and from~\eqref{eq:ICFourier}
the term $f(\Psi(\theta))$ can be seen as a function of
$\sqrt{\EE(P)}\theta$.
\begin{figure}
\centering
  \includegraphics[width=.5\linewidth]{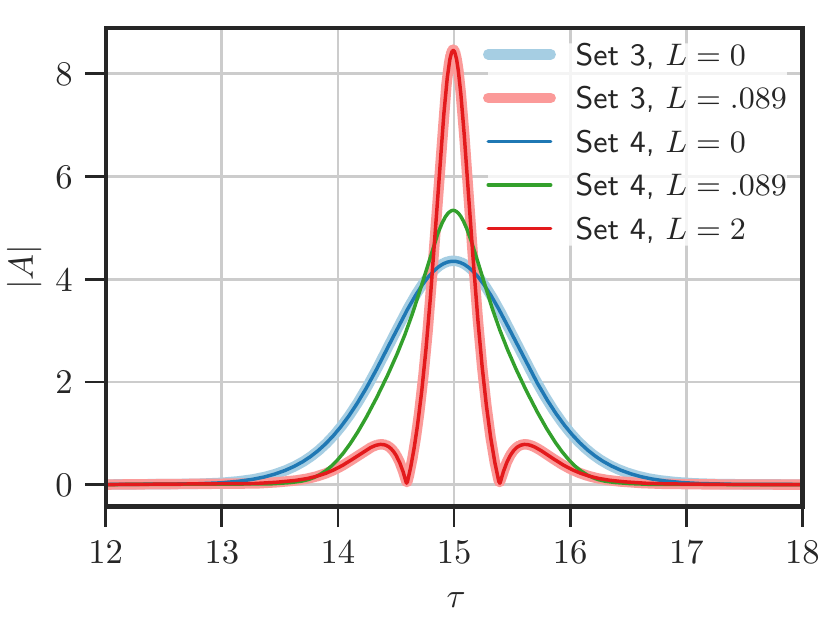}%
  \includegraphics[width=.5\linewidth]{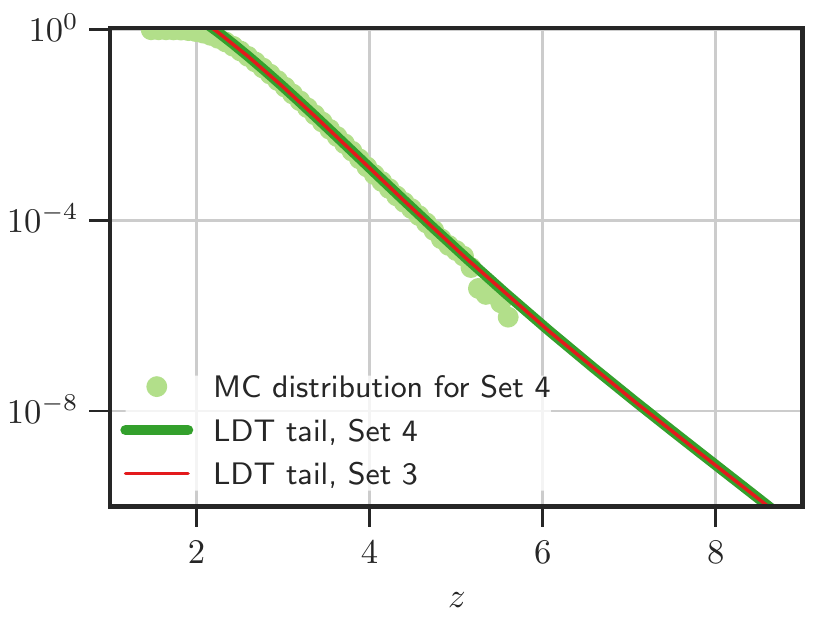}
  \caption{The two panels show how knowledge of the LDT tail at the
    output point $L$ for a given $\Delta$ and $\EE(P)$ allows us to
    recover the LDT tail for an arbitrary $\Delta'$, with the properly
    rescaled mean power $\EE(P)'$, space $L'$, and time $\tau'$. Left:
    Instanton reaching $\max_{\tau\in\Gamma}|A(L,\tau)|=8.5$, for sets
    3 and 4. Right: $\PP(\max_\tau|A(L,\tau|)\ge z$ from MC sampling
    and tail estimate for Sets 3 and 4, at an equivalent rescaled
    output point $L$. Note that not only the probability tail is the
    same for the two sets, but also the entire distribution, as the
    scale invariance establishes a complete equivalence between two
    sets having the same value of the ratio $\sqrt{\EE(P)}/\Delta$.}
\label{fig:9}
\end{figure}
\begin{itemize}[leftmargin=.1in]
\item Starting from the second property, we have that given a fixed
  spectral width $\Delta$ and a mean power $\EE(P)$, giving the cost
  function \eqref{eq:cost-waves} $E(\theta,\lambda)$, the cost
  function $E'(\theta,\lambda)$ associated to a new mean power
  $\EE(P)'$ (but same spectral width) can be written as
\begin{equation}
  E'(\theta,\lambda) = \frac{\EE (P)}{\EE (P)'} E(\theta',\lambda'),
  \quad \theta'=\theta\sqrt{\frac{\EE (P)'}{\EE (P)}}, \quad
  \lambda'=\lambda \frac{\EE (P)'}{\EE (P)}\,.
\end{equation}
Since $\lambda'$ is nothing but a rescaling of $\lambda$, and they are
both arbitrary variables, $E$ and $E'$ represent actually the same
landscape, just differing by a positive factor and a rescaling of the
variables. This implies that if we know an instanton $\theta^\star(z)$
and its associated probability $P(z)$ for the mean power $\EE(P)$, we
also know that for mean power $\EE(P)'$ the same event will have
instanton $\theta'^\star(z)=\theta^\star(z)\sqrt{\EE (P)'/\EE (P)}$
with associated probability
\begin{equation}\label{eq:LDTgen1}
  P'(z)= P(z)^{\frac{\EE(P)}{\EE(P)'}}\,.
\end{equation}
Thus, keeping $\Delta$ fixed, the LDT tails for a given $\EE(P)$ are
sufficient to generate the LDT tails for any mean power $\EE(P)'$,
using \eqref{eq:LDTgen1}.

\item Using the scale invariance of the NLSE, it is possible to make a
  similar argument to extend the LDT tails to arbitrary
  $\Delta$. Knowing that initial conditions with the same ratio
  $\sqrt{\EE(P)}/\Delta$ are scale invariant for the NLSE, one can
  pick an arbitrary spectral width $\Delta'$. This gives a new mean
  power $\EE(P)'=\EE(P)(\Delta'/\Delta)^2$, and allows us to compute
  the new length $L'=(\Delta/\Delta')^2 L$ and time coordinate
  $\tau'=(\Delta/\Delta')\tau$. Thus, a bijection is established
  between the two parameter sets, where each pair is characterized by
  the same non-dimensional instanton and same probability. Hence,
  knowing the LDT tails at different $L$ for one value of the spectral
  width, one is able to obtain the whole spatial transient of the LDT
  tails for an arbitrary spectral width. In Fig.~\ref{fig:10} the
  invariance of the non-dimensional instanton and of the LDT tail is
  shown for Sets 3 and 4, which yield the same dynamics once the
  appropriate rescaling is performed.
\end{itemize}

Figs.~\ref{fig:6} and~\ref{fig:8} confirm that the high-power pulses
arising spontaneously from a random background tend to the shape of
the PS around its maximum space-time
concentration~\cite{tikan2017universality}. Interpreting this in light
of the gradient-catastrophe
regularization~\cite{bertola2013universality}, it is clear that such
characteristic shape of the extreme power amplifications is
independent of the solitonic content of the field, although it is
shared with the local behavior of an exact solitonic solution. The
random extreme realizations quickly diverge from the PS away from the
maximum, however. In contrast, the instantons characterize all the
essential dynamics of the extreme events in integrable
turbulence. They give an approximation of the extreme excursions that
is much more accurate than the PS, as can be observed in
Fig.~\ref{fig:8}, and their shape adapts to the size of the event. In
addition, unlike the PS, they come with probabilistic information and
allows the estimation of the distribution tail, as seen in
Fig.~\ref{fig:7}, with mathematical justification in the LDT
result~\eqref{eq:13b}. Furthermore, the instantons depend on the
statistical state of the random background, as shown in
Fig.~\ref{fig:10}, while the PS is always the same. Because of these
properties and their connection with the gradient catastrophe (which
is their generating mechanism), the instantons can be important
objects for further investigations in integrable turbulence. In this
context, recent results~\cite{PNAS2018rogue} suggest that the
formation of extreme coherent structures may not necessarily be linked
to integrability, but may pertain to a more general class of systems
with instabilities (e.g.~due to non-resonant interactions) leading to
spatio-temporal concentration phenomena.

\section{Conclusions}
\label{sec:conclusions}

We have shown that tools and concepts from large deviation theory
(LDT), combined with optimization tools from optimal control, can be
used to analyze rare events in the context of dynamical systems
subject to random input in their parameters and/or their initial
conditions.  In our examples, the predictions from LDT were actually
valid in a wide region of parameter space. This means that the large
deviation regime is attained for events that are rare but still quite
frequent, and extend down to extremely low probabilities, exploring
regions unattainable through brute-force MC sampling. In addition, the
instantons provide us with information about the mechanism of the
events that can only be extracted from MC sampling via non-trivial
filtering.  Under this light, the LDT method stands as a competitive
alternative, or at least a useful complement, to brute-force MC.

\section*{Acknowledgment}

We thank Georg Stadler for useful comments regarding the optimization
method, and Gilles Francfort for suggesting the elastic rod
application. We are also grateful to Lamberto Rondoni, Themis Sapsis,
Freddy Bouchet, Hugo Touchette and Pierre Suret for interesting
discussions.

\appendix

\section{Calculations of section \ref{sec:rod}}\label{sec:app-3}
Using the convention that $\mathcal D_{N+1}=0$, the evolution equation
\eqref{eq:rod} can be rewritten as a system of first order ODEs,
\begin{equation}\label{eq:app-c6}
	 \left\{
	\begin{aligned}
	 \partial_t u_j &= p_j \\
	 \partial_t p_j &= \frac{\mD_{j+1}}{\Delta x^2}(u_{j+1}-u_j) - \frac{\mD_j}{\Delta x^2}(u_j - u_{j-1}) + \delta_{j,N}\frac{r(t)}{\Delta x}
  	\end{aligned}\right. \,,\quad j=1,...,N\,
\end{equation}
with fixed boundary condition in the origin,
\begin{equation}\label{eq:app-c3}
	u_0(t) = 0 \,,
\end{equation}
and initial conditions
\begin{equation}\label{eq:app-c4}
	u_j(0) = 0, \quad p_j(0) = 0\,.
\end{equation}
To make the notation compact, we will use:
\begin{equation}
	X= \left(\begin{array}{c}u\\p\end{array}\right), \quad Y =\left(\begin{array}{c}\eta\\\mu\end{array}\right),
\end{equation}
column vectors in $\mathbb{R}^{2N}$.
Then, \eqref{eq:app-c6} can be written as
\begin{equation}\label{eq:app-c7}
	\partial_t X = b(X,\theta)\,,
\end{equation}
where $b(X,\theta)$ is the $2N$-dimensional vector with the components of the RHS of \eqref{eq:app-c6}. Note that \eqref{eq:app-c7} is in the general form \eqref{eq:detdyn00} (linear system of ODEs), and this is helpful to make direct contact with the formulas \eqref{eq:adjoint} and \eqref{eq:gradE3}, and thereby compute the gradient of the cost function \eqref{eq:Ediscr} as
\begin{equation}\label{eq:app-c1}
	\nabla_\theta E = \nabla_\theta I(\theta) -\int_0^T (\partial_\theta b)^T Y \, dt\,,
\end{equation}
with $Y$ the adjoint field to $X$. Let us start by deriving the adjoint equation.
One can easily check that the linearization of the operator $b(X,\theta)$ for small variations of $X$ reads
\begin{equation}\label{eq:app-c8}
	\begin{aligned}
		&\partial_X b(\theta) = \left(\begin{array}{cc}0 & \text{Id} \\ B(\theta) & 0\end{array}\right) , \\
		\text{with} \quad  B_{jk} = & \frac{\mathcal D_{j+1}}{\Delta x^2}(\delta_{j+1,k}-\delta_{j,k}) - \frac{\mathcal D_j}{\Delta x^2}(\delta_{j,k}-\delta_{j-1,k}).
	\end{aligned}
\end{equation}
$\text{Id}$ is the $N\times N$ identity matrix and we recall that $\mathcal D_j=\mathcal{D}(\theta_j)$, by \eqref{eq:PDFchain}.
It is the adjoint operator $(\partial_X b)^T$ that we need to compute, defined implicitly by the identity
\begin{equation}\label{eq:app-c10}
	\left\langle(\partial_X b)^TY,X'\right\rangle_{\mathbb{R}^{2N}}=\left\langle Y,\partial_X b\, X'\right\rangle_{\mathbb{R}^{2N}}\,,
\end{equation}
where $\<\cdot,\cdot\>_{\mathbb{R}^{2N}}$ denotes the standard scalar product in $\mathbb{R}^{2N}$.
Using \eqref{eq:app-c10} we obtain,
\begin{equation}\label{eq:app-c6a}
	\begin{aligned}
	\left\langle Y,\partial_X b\, X'\right\rangle_{\mathbb{R}^{2N}}&= \sum_{j=1}^{N} \left(  \eta_j p'_j + \mu_j \left( \frac{\mathcal D_{j+1}}{\Delta x^2}(u'_{j+1} - u'_j) - \frac{\mathcal D_{j}}{\Delta x^2} ( u'_j - u'_{j-1} ) \right)  \right) \\
	& = \sum_{j=1}^{N} \left(  \eta_j p'_j + \left( \frac{\mathcal D_{j+1}}{\Delta x^2}(\mu_{j+1} - \mu_j) - \frac{\mathcal D_{j}}{\Delta x^2} ( \mu_j - \mu_{j-1} ) \right) u'_j  \right)\,,
	\end{aligned}
\end{equation}
where in the last passage we just reorganized the indices in the sum in an equivalent way, provided that we assume the boundary condition
\begin{equation}\label{eq:app-c9a}
	\mu_0(t) = 0 \,.
\end{equation}
Comparing the last line of \eqref{eq:app-c6a} with the LHS of \eqref{eq:app-c10}, we deduce that
\begin{equation}\label{eq:app-c10a}
	(\partial_X b)^T = \left(\begin{array}{cc}0 & B(\theta) \\ \text{Id} & 0\end{array}\right)
\end{equation}
which is the transpose of the RHS of \eqref{eq:app-c8} ($B(\theta)$ is symmetric), as we should expect. Though, starting from the identity \eqref{eq:app-c10} is the rigorous way to obtain the adjoint operator, making the proper boundary conditions arise naturally.
Plugging the result \eqref{eq:app-c10a} into \eqref{eq:adjoint}, we finally obtain the adjoint equation
\begin{equation}\label{eq:app-c6b}
	 \left\{
	\begin{aligned}
	 \partial_t \eta_j &= \frac{\mD_{j+1}}{\Delta x^2}(\mu_{j+1}-\mu_j) - \frac{\mD_j}{\Delta x^2}(\mu_j - \mu_{j-1}) \\
	 \partial_t \mu_j &=  \eta_j
  	\end{aligned}\right. \,,\quad j=1,...,N\,,
\end{equation}
with boundary condition \eqref{eq:app-c9a}. To obtain the correct conditions at final time, it is sufficient to observe that the final conditions of \eqref{eq:adjoint} now read
\begin{equation}
  \label{eq:adjoint-app}
  \eta_j(T) = \lambda \partial_{u_j} f(u(T)) = \lambda \delta_{j,N} \,, \qquad \mu_j(T) =0.
\end{equation}

Let us now compute $(\partial_\theta b)^T$, again starting from the definition of the adjoint operator:
\begin{equation}\label{eq:app-c11a}
	\left\langle(\partial_\theta b)^TY,w\right\rangle_{\mathbb{R}^{N}}=\left\langle Y,\partial_\theta b\, w\right\rangle_{\mathbb{R}^{2N}},
\end{equation}
where $w\in\mathbb{R}^N$ and
\begin{equation}
	\begin{aligned}
	(\partial_\theta b) &= \left(\begin{array}{c}0 \\ \nabla_\theta B(\theta) \end{array}\right) 	\quad \text{(two } N\times N \text{ blocks)}\\
	\quad (\nabla_\theta B)_{jk}= \frac{\mathcal{D}'(\theta_{j+1})}{\Delta x^2}&(u_{j+1}-u_j) \delta_{j+1,k} - \frac{\mathcal{D}'(\theta_{j})}{\Delta x^2}(u_{j}-u_{j-1}) \delta_{j,k}.
	\end{aligned}
\end{equation}
With the convention that $\mathcal{D}'(\theta_{N+1}=0)$, a straightforward calculation yields
\begin{equation}\label{eq:app-c78}
	\begin{aligned}
	\left\langle Y,\partial_\theta b\, w\right\rangle_{\mathbb{R}^{2N}}&= \sum_{j=1}^{N} \mu_j \left(  \frac{\mathcal{D}'(\theta_{j+1})}{\Delta x^2} (u_{j+1}-u_j)w_{j+1} - \frac{\mathcal{D}'(\theta_{j})}{\Delta x^2} (u_{j}-u_{j-1})w_{j} \right) \\
	& =  \sum_{j=1}^{N} \left(  \frac{\mathcal{D}'(\theta_{j})}{\Delta x^2} (u_{j}-u_{j-1}) (\mu_{j}-\mu_{j-1}) \right) w_{j}\,,
	\end{aligned}
\end{equation}
from which, comparing with the LHS of~\eqref{eq:app-c11a}, we observe that
\begin{equation}
	((\partial_\theta b)^T Y)_j = \mathcal{D}'(\theta_{j})\frac{u_{j}-u_{j-1}}{\Delta x}\frac{\mu_{j}-\mu_{j-1}}{\Delta x}\,.
\end{equation}
Now, integrating in time according to \eqref{eq:app-c1}, 
\begin{equation}
	\int_0^T((\partial_\theta b)^T Y)_j dt= \mathcal{D}'(\theta_{j}) \int_0^T\frac{u_{j}-u_{j-1}}{\Delta x}\frac{\mu_{j}-\mu_{j-1}}{\Delta x} dt\,,
\end{equation}
leads to~\eqref{eq:gradRod1}.
%
%
\bibliographystyle{abbrv}
\bibliography{bib}
\end{document}